\documentclass[fleqn,usenatbib]{mnras}

\usepackage{newtxtext,newtxmath}
\usepackage{xcolor}
\usepackage{soul}

\usepackage[T1]{fontenc}

\DeclareRobustCommand{\VAN}[3]{#3}
\let\VANthebibliography\thebibliography
\def\thebibliography{\DeclareRobustCommand{\VAN}[3]{##3}\VANthebibliography}

\defcitealias{Noor_2025}{Paper~I} 
\usepackage{float}
\usepackage{ragged2e} 
\usepackage{gensymb}
\usepackage{graphicx}	
\usepackage{amsmath}	
\usepackage{lineno}

\title[Dynamically Active Planetary Debris Disks II]{Activity in White Dwarf Debris Disks II: {\em WISE} Mission Traces a Decade of Disk Evolution and Persistence}

\author[H.~T.~Noor et al.]{
Hiba Tu Noor,$^{1}$\thanks{E-mail: hiba.noor.19@ucl.ac.uk} Jay Farihi,$^{1}$
Scott J. Kenyon$^{2}$
\\
$^{1}$Department of Physics and Astronomy, University College London, London WC1E 6BT, UK\\
$^{2}$Smithsonian Astrophysical Observatory, Cambridge MA 02138, USA}
\date{Accepted XXX. Received YYY; in original form ZZZ}

\pubyear{2024}

\begin{document}
\label{firstpage}
\pagerange{\pageref{firstpage}--\pageref{lastpage}}
\maketitle

\begin{abstract}
This study presents multi-epoch photometry from the {\em Wide-field Infrared Survey Explorer} to investigate variability in white dwarf debris disks, utilizing the full 14.5\,yr mission datasets pre- and post-cryogen. Unambiguous variations are detected across the bulk of the 52 confirmed dusty white dwarfs and are consistent with universal variability after correcting for the sensitivity of {\em WISE} relative to {\em Spitzer}.  Some disk light curves exhibit trends such as possible oscillations, sudden brightening, and steadily decaying fluxes, but all infrared excesses persist utterly. On baselines longer than 0.5\,yr, flux changes are significantly skewed toward decreases, in contrast to the symmetric distribution found at shorter baselines with {\em Spitzer}, consistent with gradual dust depletion over time. Disk emission generally becomes redder when dimmer and bluer when brighter, consistent with changes in the grain size distribution following collisions, although five disks exhibit the opposite trend possibly due to dust production near apastron.  A moderate positive correlation is observed between disk luminosity and photospheric Ca abundance, consistent with accretion from an optically thin disk component. Taken together with the {\em Spitzer} legacy of dusty white dwarfs, these results provide a comprehensive view of debris disk variability and provide empirical constraints on models of disk structure and evolution.

\end{abstract}

\begin{keywords}
circumstellar matter -- planetary systems -- white dwarfs
\end{keywords}



\section{Introduction}

Post-main sequence planetary systems often manifest via the deposition of metals in white dwarf atmospheres. Given their high surface gravities, heavy elements are expected to rapidly sink below the photosphere on timescales of days to Myr in stars sufficiently cool that radiative levitation is negligible \citep{Paquette_1986, Koester_2009}, where these timescales are always orders of magnitude shorter than the corresponding white dwarf cooling ages. The detection of atmospheric metals in up to 50\,per cent of typical, single white dwarfs \citep{Zuckerman_2010, Koester_2014, OuldRouis_2024} thus requires an external source.

Myriad evidence demonstrates these atmospheric metals have a planetary origin. In the consensus model, planetary bodies are perturbed onto star-grazing orbits that pass within the stellar Roche radius, where they are tidally disrupted and subsequently accreted \citep{Jura_2003}. This interpretation is supported by disks of circumstellar dust and gas, detected and confirmed around over 60 white dwarfs \citep[e.g.\ ][]{Gansicke_2006, Jura_2007, vonHippel_2007, Farihi_2009, Girven_2012, Melis_2020}, with more than 100 additional candidates identified through potential infrared excesses \citep{Dennihy_2017, RebassaMansergas_2019, MadurgaFavieres_2024}. Over a dozen white dwarfs also exhibit transits attributed to planetary debris, further corroborating this picture whereby planetary systems survive into the post-main sequence \citep[e.g.][]{Guidry_2021,Bhattacharjee_2025}.

While the presence of planetary debris around white dwarfs is well established, the physical structure and dynamical evolution of these disks remain empirically challenging to constrain.  Infrared excesses are predominantly detected in the range 2--5\,$\upmu$m, largely reflecting the wavelength coverage and sensitivity of past and present surveys, tracing dust at temperatures of approximately 1000\,K, located near the Roche limit, typically at $\sim 0.1\,R_{\odot}$ \citep[for a review, see][]{Farihi_2016}.  Mid-infrared variability is ubiquitous among these systems, with fluctuations in brightness of tens of per cent in amplitude \citep[e.g.\ ][]{Xu_2018b, Swan_2019, Guidry_2024} and consistent with ongoing collisions of planetesimal debris, which continually modify the dust population through both production and destruction \citep[e.g.\ ][]{Farihi_2018, Swan_2020, Noor_2025}. 

Such variability is fundamentally at odds with the previously-accepted and pioneering model of a geometrically thin, optically thick ring \citep{Jura_2003, Rafikov_2011}, which predicts a passively evolving, quiescent disk that is incapable of accounting for both the frequency and spectral properties of the observed changes \citep{Noor_2025}. Additional observations are similarly incompatible with a flat and opaque disk, including transiting debris systems that require occulting structures that are hundreds of km in vertical extent \citep{Vanderburg_2015, Vanderbosch_2021, Farihi_2022}, and cases where disk luminosities exceed the maximum allowed by the model \citep{Jura_2007b, Dennihy_2017}. An alternative description of these disks is thus required, with community consensus and stronger theoretical and empirical support.

Developing the necessary models requires empirical constraints on both disk structure and evolution, and mid-infrared monitoring provides a direct means of obtaining them by tracing changes in the emitting dust population. However, observations of dusty white dwarfs are challenging and generally require space-based facilities; with such resources limited, it is therefore paramount to thoroughly exploit available datasets, particularly those with the longest observing baselines. Previous studies of disk variability have relied predominantly on {\em Spitzer} \citep{Swan_2020}, while {\em WISE} observations have largely been restricted to individual systems \citep{XuJura_2012, Farihi_2018} or subsets of the known disk population over partial mission coverage \citep{Swan_2019, Guidry_2024}, leaving the full population uncharacterized. With the {\em WISE} mission now complete, maximum scientific return can be achieved by analyzing all bona fide and candidate dusty white dwarfs across the entire 14.5\,yr archive.

The present study is the second of a two-part investigation into the variability of mid-infrared emission from white dwarf debris disks. The first part analyzed all available multi-epoch photometry of known white dwarf disks observed by {\em Spitzer}, capturing variability on timescales from minutes to weeks, with sparse baseline coverage extending to 15\,yr for objects observed during the cryogenic mission \citep[][hereafter Paper~I]{Noor_2025}. This work extends the investigation using the {\em Wide-field Infrared Survey Explorer} ({\em WISE}; \citealt{Wright_2010}), which provides all-sky coverage and a roughly uniform six month cadence over a 14.5\,yr baseline, probing a temporal regime largely inaccessible to {\em Spitzer}.

This work is organized as follows. Section~\ref{sec:sample_selection} describes the sample and data selection. Section~\ref{sec:results} details the criteria used to establish variability and presents the results of the survey, discussing mid-infrared variability both at the population level and for individual systems. Section~\ref{sec:discussion} discusses the implications of these findings and a summary is given in Section~\ref{sec:conclusion}. Light curves for all well-sampled targets are included in the Appendix.

\section{Sample selection and processing}
\label{sec:sample_selection}

This section describes the sample selection, {\em WISE} data retrieval, and photometric filtering process.

{\em WISE} initially surveyed the entire sky in four bands: W1 (3.4\,$\upmu$m), W2 (4.6\,$\upmu$m), W3 (12\,$\upmu$m), and W4 (22\,$\upmu$m). Following depletion of the cryogen, only W1 and W2 remained operational during the NEOWISE reactivation mission \citep{Mainzer_2014}. The satellite had a 1.5\,h orbit and typically observed a given source 10--20 times during a single epoch, with observations repeated roughly every 0.5\,yr. With its final data release in 2024 November, the mission provides a 14.5\,yr baseline for each target and enables variability to be probed on timescales of months to years. 

The primary science sample is drawn from 64 metal-polluted white dwarfs with unambiguous infrared excesses consistent with $T \approx 1000\,$K circumstellar dust, for which source confusion can be confidently ruled out. Most of these disks were first identified through {\em Spitzer} \citep[e.g.\ ][]{Jura_2007, Farihi_2009, Girven_2012, Brinkworth_2012}, with additional detections from ground-based observations \citep{Becklin_2005, Melis_2011}, and {\em JWST} \citep{Farihi_2025}. All host white dwarfs in this sample exhibit atmospheric metal pollution and, where mid-infrared spectroscopy is available, all show solid-state emission from micron-sized silicate grains \citep{Reach_2005b, Jura_2007b, Jura_2009, Farihi_2025}. This sample thus represents a bona fide population of dusty white dwarfs and is hereafter referred to as Sample~I.

A second sample of disk candidates is constructed using published catalogues of {\em WISE}-selected infrared excesses \citep{Dennihy_2017, RebassaMansergas_2019, Xu_2020, Lai_2021}. These targets are analyzed separately owing to their uncertain nature and the risk of source confusion in {\em WISE} photometry \citep{Melis_2011, Hoard_2013, Dennihy_2020a}. On average, these candidate disks are fainter than Sample~I by approximately 0.8\,mag in W1 (roughly a factor of two). This sample is hereafter referred to as Sample~II.

The coordinates of each star in both samples are used to query the ALLWISE Multiepoch Photometry Table and the NEOWISE-R Single Exposure Source Table, employing a 5\,arcsec search radius, increased to 10\,arcsec where necessary to account for proper motion. Measurements are retrieved in the W1 and W2 bands, and several quality cuts are applied to the {\em WISE} parameters to filter the observations, as outlined below.

Data are selected to minimize photometric contamination from proximity to image artefacts, the South Atlantic Anomaly, and scattered moonlight, requiring \textsc{cc\_flags} $=0$, \textsc{saa\_sep} $>5$, and \textsc{moon\_masked} $=0$, which typically removes less than 5--10\,per cent of light curve data points. Observations with degraded image quality, including smeared point-spread functions (PSFs) or spurious detections, are further excluded by requiring \textsc{qual\_frame} $>5$ and \textsc{qi\_fact} $>0.5$. For targets located in moderately crowded fields (typically, Galactic latitude $|b|\lesssim20\degr$), reliable photometry is retained using \textsc{nb} $>1$ and \textsc{na} $=1$, corresponding to cases where multiple PSF components are fitted and sources are actively de-blended. The ALLWISE Reject Table includes entries for objects with signal-to-noise (S/N) below the catalogue limit, duplicate measurements from the source catalogue, and spurious detections resulting from noise or image artefacts; observations listed in this catalogue and measurements flagged as upper limits are discarded. Finally, photometric contamination is checked for each star using neighbouring sources resolved by {\em WISE}: objects with $\Delta m\leq1$\, mag and within 7.8\,arcsec of the target are flagged \citep[this is $1.3\times$ the W1 PSF full width at half maximum, the critical distance for resolving overlapping sources;][]{Dennihy_2020a}. Variability observed in these sources should be treated with caution, as photometric contamination may dilute any intrinsic flux changes, but more importantly, can introduce a time-variable component owing to varying degrees of de-blending fidelity from changes in scanning direction and especially proper motion of the science targets.

Only objects with two or more observational epochs are retained, requiring a minimum of six individual measurements per epoch to pass the above quality criteria, ensuring the weighted mean flux and its uncertainty are well determined at each epoch. After applying these cuts, 52 and 59 white dwarfs remain in Samples~I and II, respectively, and are listed in Tables~\ref{tab:known disks} and \ref{tab:candidate disks}.

\begin{table}
\begin{center}	 
\caption{Results for Sample~I, confirmed dusty white dwarfs.}
\label{tab:known disks}

\begin{tabular}{lllllr}

\hline
\hline
WD      & 
\multicolumn{2}{c}{|$\Delta F_{3.4, \mathrm{max}}$|}  &
WD      & 
\multicolumn{2}{c}{|$\Delta F_{3.4, \mathrm{max}}$|}  
\\

&
$\%$ &
$\upsigma$&
&
$\%$ &
$\upsigma$
\\

\hline

\multicolumn{6}{c}{Isolated sources}  \\ 

\multicolumn{3}{c}{Dust}  &
\multicolumn{3}{c}{Dust and Ca \textsc{ii} emission}\\

\hline

0106$-$328      &41             &3.7    &
0145+234        &360            &75\\

0110$-$565      &41             &6.0    &
J0347+1624      &62             &8.6\\

0146+187        &36             &6.2    &
J0510+2315      &230            &22\\

J0207+3331      &18             &3.7   &
J0529$-$3401    &30             &5.9\\

0300$-$013      &41             &5.2    &
0842+231        &32             &4.1\\

0307+077        &170            &13     &
0842+572        &19             &21 \\

0408$-$041      &26             &14     &
1116+026        &14             &2.6\\

0420$-$731      &12             &4.3    &
1145+288        &99             &4.7\\

J0537$-$4758    &28             &3.3    &
1226+110        &34             &5.0\\

J0707$-$7438    &29             &2.8    &
1349$-$230      &76             &2.9\\

0716+404        &88             &4.6    &
1622+587        &41             &4.1\\

J0802+5631      &26             &3.1    &
J1930$-$5028    &120            &5.6\\

0843+516        &32             &2.7    &
2212$-$135      &42             &2.4\\

1015+161        &38             &4.3    \\

1150$-$153      &19             &5.0\\
1225$-$079      &76             &3.0\\
1232+563        &84             &7.2\\
1457$-$086      &190            &7.2\\
1536+520        &22             &4.9\\
1541+650        &16             &5.1\\
1612+554        &42             &4.3\\
J1617+1620      &120            &6.1\\
2115$-$560      &21             &6.1\\
2132+096        &800$^{\rm a}$  &5.3\\
2207+121        &79             &4.2\\
2221$-$165      &110            &4.9\\
2326+049        &17             &12 \\
2329+407        &22             &6.6\\

\hline
\multicolumn{6}{c}{With neighbour(s)}  \\ 

\multicolumn{3}{c}{Dust}  &
\multicolumn{3}{c}{Dust and Ca \textsc{ii} emission}\\
\hline

0435+410        &29             &3.8    &
J0006+2858      &19             &1.7\\

1018+410        &120            &4.1    &
J0234$-$0406    &55             &3.3\\

J1221+1245      &70             &8.0    &
J0644$-$0352    &110            &4.3\\

1729+371        &14             &2.9    &
J2100+2122      &189            &21\\

J1931+0117      &22             &9.3    &
2133+242        &92             &16\\

\hline
\end{tabular}
\end{center}
\justifying
\noindent
{\em References:}  
\citet{ZuckermanBecklin_1987}; \citet{Becklin_2005};  \citet{Gansicke_2006}; \citet{Kilic_2006}; \citet{Jura_2007}; \citet{KilicRedfield_2007}; \citet{Mullally_2007}; \citet{Gansicke_2008};  \citet{Brinkworth_2009}; \citet{Farihi_2009}; \citet{Farihi_2010}; \citet{Farihi_2011}; \citet{Debes_2011};  \citet{Melis_2011};  \citet{Farihi_2012}; \citet{Girven_2012}; \citet{XuJura_2012}; \citet{Kilic_2012};   \citet{Brinkworth_2012}; \citet{Hoard_2013}; \citet{Wilson_2014};   \citet{Bergfors_2014};  \citet{Rocchetto_2015}; \citet{RebassaMansergas_2019};  \citet{Dennihy_2020a};  \citet{Melis_2020};  \citet{GentileFusillo_2021b};  \citet{Lai_2021}; \citet{Farihi_2025}; \citet{LeBourdais_2025}.\\
\noindent {\em Notes}: $^{\rm a}$ Transition between epoch with excess below the detection threshold and one with a significant excess gives large percentage change.
\end{table}

\begin{table}
\begin{center}	 
\caption{Results for Sample~II, candidate dusty white dwarfs.}
\label{tab:candidate disks}

\begin{tabular}{lllllr}

\hline
\hline
WD      & 
\multicolumn{2}{c}{|$\Delta F_{3.4, \mathrm{max}}$|}  &
WD      & 
\multicolumn{2}{c}{|$\Delta F_{3.4, \mathrm{max}}$|}  
\\

&
$\%$ &
$\upsigma$&
&
$\%$ &
$\upsigma$
\\

\hline

\multicolumn{6}{c}{Isolated sources }  \\ 

\hline

0107-192        &41             &3.7    &
J0412$-$4510    &13             &4.5 \\

J0119+1044      &100            &5.0    &
J0603+4518      &14             &2.4 \\

0253+508        &20             &2.3    &
J2340$-$3708    &31             &3.2 \\

\hline
\multicolumn{6}{c}{With neighbour(s)} \\

\hline

J0050$-$0326    &27             &3.2    &
J1154$-$3101    &22             &4.5\\  

J0052+4505      &31             &3.1    &  
J1216+7455      &10             &2.7\\     

J0115$-$5207    &54             &3.5    &
J1305+1525      &55             &1.9\\  	 

J0119$-$7655    &58             &6.7    &  
1318+648        &39             &4.9\\     

J0125+1811      &23             &3.5    &  
J1322$-$1210    &38             &2.0\\     

J0205$-$7941    &41             &6.4    &  
1352+013        &20             &2.0\\     

J0329$-$4738    &23             &4.3    &  
1426+442        &21             &3.0\\     

J0518+6753      &22             &4.1    &  
J1509+1411      &31             &3.4\\     

J0529$-$4303    &31             &3.6    &  
J1539$-$3910    &60             &3.7\\     

J0649$-$7624    &21             &2.8    &  
1612+552        &17             &3.4\\     

J0701+2321      &50             &5.6    &  
1721+050        &33             &3.8\\    

J0702+0003      &22             &4.2    &  
J1728+2053      &52             &5.5\\     

J0723+6301      &84             &5.3    &  
J1806+2731      &26             &2.3\\     

J0730+2716      &28             &4.1    &  
J1814$-$7355    &17             &7.8\\     

J0731+2417      &30             &3.8    &  
J1903+6035      &7.3            &4.7\\     

J0734$-$6011    &18             &3.6    &  
J1939+0932      &6.7            &2.0\\     

0823+819        &23             &3.5    &  
J1949+7007      &23             &3.0\\     

0841+336        &76             &3.8    &  
J2008$-$6604    &18             &7.7\\     

J0854$-$7646    &64             &4.7    &  
J2158$-$5853    &19             &4.4\\     

J0924$-$2423    &30             &3.1    &  
J2205$-$4610    &39             &2.5\\     

J0951+0749      &50             &2.0    &  
J2223$-$2510    &33             &2.9\\     

J1017$-$3236    &14             &2.9    &  
2231$-$387      &17             &3.3\\     

J1030$-$1435    &55             &4.0    &  
J2253+0833      &31             &2.6\\     

J1039$-$0325    &42             &6.7    &  
J2305+5125      &22             &3.5\\     

J1055$-$0237    &28             &7.1    &  
J2316$-$5529    &42             &4.3\\     

1122+426        &17             &2.9    &  
J2330+2934      &98             &6.4\\     

J1146$-$3636    &31             &2.3 \\

\hline
\end{tabular}
\end{center}
\justifying
\noindent
{\em References:} \citet{Dennihy_2017}; \citet{RebassaMansergas_2019}; \citet{Dennihy_2020a}; \citet{Lai_2021}.
\end{table}

{\em WISE} reported Vega magnitudes are retrieved at each epoch and converted to flux densities \citep{Wright_2010} for all stars in both samples. For each white dwarf, the photospheric contribution is determined using grids of pure hydrogen or helium atmosphere models spanning effective temperatures 6000--30\,000\,K \citep{Koester_2010}, selected according to the published spectral classification of the host star. The models are fitted to optical photometry from surveys such as Pan-STARRS and SkyMapper \citep{Chambers_2016, Wolf_2018} by $\upchi^2$ minimization, with the resulting effective temperatures found to be broadly consistent with literature values compiled in the Montreal White Dwarf Database\footnote{\url{https://montrealwhitedwarfdatabase.org/}}. The stellar flux, extrapolated to the W1 and W2 bandpasses using synthetic photometry, is subtracted from the total measured flux at each epoch to isolate the dust emission. This excess is fitted with a single-temperature blackbody, yielding temperature $T_{\rm IR}$, dust orbital radius $R_{\rm IR}$, and disk luminosity $L_{\rm IR}$. The resulting W1 normalized light curves for all targets are presented in Appendix~\ref{sec: light curves}.

To assess the significance of any variability associated with circumstellar dust, a control sample is assembled, consisting of 1100 white dwarfs with no previously reported infrared excesses or variability, hereafter referred to as Sample~III. This sample is drawn from high-confidence {\em Gaia} EDR3 white dwarfs \citep{GentileFusillo_2021} with $P_{\rm WD}>0.99$, hydrogen-dominated atmospheres ($\upchi_{\rm H}< \upchi_{\rm He}$ or $\upchi_{\rm mixed}$), $T_{\rm eff}>13\,000$\,K, and $M=0.58$--0.62\,$M_{\odot}$, selected to avoid ZZ~Ceti pulsators and unresolved binaries, respectively. The sample is selected solely based on these stellar properties, with no criteria on sky position or local source density. All 1100 objects are widely distributed across the sky, spanning the full ranges of right ascension (0.03\degree to 359.92\degree) and declination ($-$88.1\degree to +87.8\degree) with a median nearest neighbour separation of 1.13\degree. While field-dependent effects, e.g.\ source confusion, are observed in {\em WISE} photometry, this broad sky distribution ensures that any potential local field effects are minimal. The {\em WISE} data for these objects are processed using the same filtering and quality cuts as applied to Samples~I and II, providing a consistent baseline for comparison.

\section{Results and Analysis}
\label{sec:results}

This section describes the criteria used to identify significant variability in the target samples. Population-wide trends are examined separately for Samples~I and II, and the behaviour of Sample~II is evaluated in the context of potential source confusion and the resulting photometric contamination. The section concludes with a discussion of selected light curves.

\subsection{The variability fractions of Sample~I, II, and III}
\label{sec:establishing var}

For each source, W1 and W2 light curves are constructed using the weighted mean flux and its associated uncertainty (the standard error of the mean) of the individual flux measurements within a given {\em WISE} epoch.  Fractional changes between epochs are defined as $\Delta F=(f_i - f_{j})/f_i$, and their formal significance as 
$\upsigma^2 = (f_i - f_{j})^2/(\upsigma_i^2+\upsigma_{j}^2)$. 

\begin{figure}
\includegraphics[width=\columnwidth]{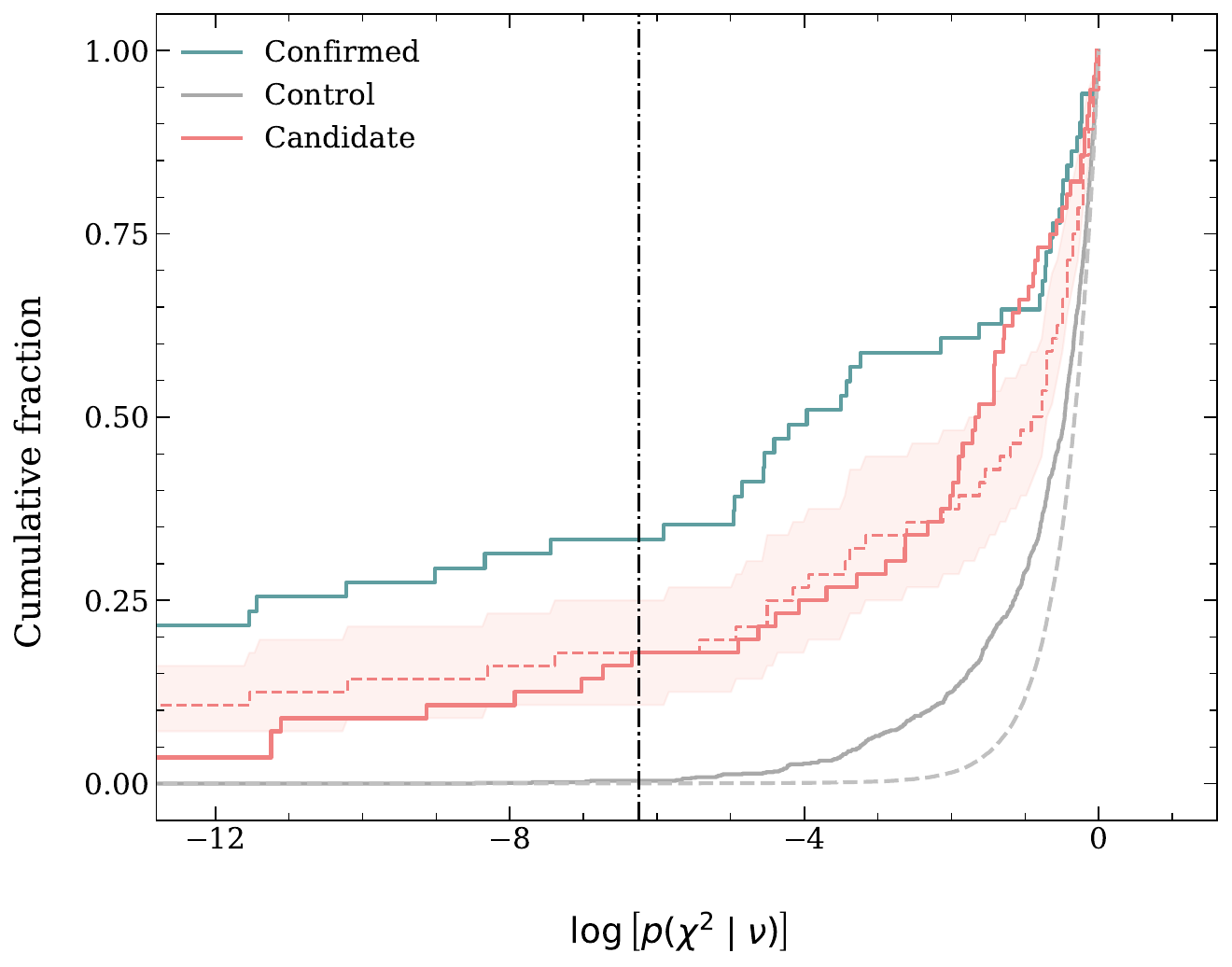}
\caption{Cumulative distribution functions of the probabilities for a constant-flux model for W1 light curves.  Solid lines show Sample~I in teal, Sample~II in coral, and Sample~III in grey. The  dash–dotted vertical line marks the probability corresponding to a $5\upsigma$ significance threshold, and lower values indicate stronger deviations from the flat light curve model. Sample~I exhibits the largest fraction of sources consistent with genuine variability.  The grey dashed line shows the CDF expected for constant sources subject to quasi-Gaussian photometric noise in their light curves.  The dashed coral line and shaded region show the CDF and envelope derived from bootstrap realizations of a mixture model used to reproduce the Sample~II distribution from Samples~I and III (see Section~\ref{sec:establishing var} for details). }
\label{fig:establish_var}
\end{figure}

To quantify variability over the full mission duration, each target is tested against a constant flux model using the $\upchi^2$ statistic, defined as $\upchi^2 = \sum_{i=1}^{N} \left(\frac{f_i-\bar{f}}{\sigma_i}\right)^2$, where $f_i$ and $\upsigma_i$ are the flux and associated uncertainty at epoch $i$, and $\bar{f}$ is the weighted mean flux over $N$ epochs. From this, the corresponding probability $p(\upchi^2|\upnu)$ per degree of freedom $\upnu$ is calculated. In this metric, the smaller the $p$-value, the less likely that the observed light curve fluxes can be accounted for by the measurement errors alone, and thus increasingly negative $\log[p(\upchi^2|\upnu)]$ implies a greater likelihood of genuine variability. Figure~\ref{fig:establish_var} shows the cumulative distribution functions (CDFs) of $\log[p(\upchi^2|\upnu)]$ for all three samples, as derived from W1 light curves (used here owing to lower S/N in W2).  Sample~I exhibits the largest fraction of sources at low $p$-values, indicating the highest fraction of variable sources. Adopting a conventional 5$\upsigma$ threshold where $p=5.7 \times10^{-7}$ for a normal distribution, 0.4\,per cent of Sample~III stars have $p$-values below this value. Although a small fraction, at face value 0.004 is inconsistent with $5.7\times10^{-7}$ as expected for random noise alone. This inconsistency likely reflects limitations of the data and the statistical assumptions underlying the $\upchi^2$ test. One likely contributor is that the $\upchi^2$ test assumes independent measurements, whereas {\em WISE} photometry exhibits correlated systematics within observing visits \citep{Cutri_2012}; such correlations reduce the effective degrees of freedom relative to the assumed $\upnu$, inflating the $\upchi^2$ statistic, and thereby producing an excess of smaller $p$-values.

To further assess this outcome, synthetic light curves are generated as follows. A star is drawn randomly from Sample~III, its median absolute flux is adopted as a constant across all epochs, and the associated uncertainties are retained. At each epoch, Gaussian noise is added with a standard deviation equal to the corresponding epoch error. Note that since these uncertainties include calibration contributions, the simulated noise is only approximately Gaussian. The $\upchi^2$ statistic and corresponding $p(\upchi^2|\upnu)$ are computed as before. Repeating this process over 50\,000 iterations yields a simulated probability distribution for non-variable sources observed with the same cadence and noise properties as the control stars. The resulting CDF is shown as a grey dashed curve in Figure~\ref{fig:establish_var}. At the lowest $p$-values, the simulated distribution closely follows that of Sample~III, but they diverge near $p=10^{-5}$, indicating a modest excess in the control sample relative to the simulations.  These results are highly unlikely to reflect any real, low-level variability among the Sample~III stars, but instead indicate systematics such as fluctuations in instrument calibration, including array degradation over time, or more likely, (time-dependent) source confusion within the {\em WISE} photometric pipeline.  The Sample~III stars are not filtered for neighbouring source density and should thus suffer from source confusion as often as any field star of similar brightness.  Therefore, the CDF of Sample~III may be a good indicator of the effects of photometric contamination on W1 light curves of white dwarfs.

The  $5\upsigma$ variability fraction of Sample~I is 33\,per cent, and clearly lower than the universal variability inferred from a complete {\em Spitzer} legacy study of white dwarf debris disks \citepalias{Noor_2025}.  This would remain true even if the threshold is lowered to $3\upsigma$, so that sources with $p$-values lower than 0.0027 are considered variable. Such a discrepancy is expected given the higher sensitivity of {\em Spitzer}, for which the median fractional photometric uncertainty is approximately 0.7\,per cent compared to 7\,per cent for {\em WISE} across the targets sampled by both facilities.  {\em Spitzer} thus enabled smaller absolute changes in dust emission around the same objects to be more readily detected, especially with differential photometry using field stars.  

To quantify the effective difference in sensitivity of the two satellites as applied to dusty white dwarfs, a Monte Carlo analysis is performed comparing the distribution of fractional flux changes measured with {\em Spitzer} in IRAC channel 1 (3.6\,$\upmu$m) to the distribution of fractional photometric uncertainties in W1 (3.4\,$\upmu$m). In each trial, a fractional variability amplitude $\Delta F$ is randomly drawn from the observed {\em Spitzer} distribution and compared to a fractional uncertainty drawn from the {\em WISE} error distribution; a flux change is considered detectable if the drawn {\em Spitzer} amplitude exceeds three times the corresponding {\em WISE} uncertainty.  Repeating this procedure over 100\,000 trials yields a W1 detection efficiency of approximately 24\,per cent relative to {\em Spitzer}. Accounting for this sensitivity difference, the 33\,per cent variability fraction measured for Sample~I in {\em WISE} is consistent with ubiquitous variability, as inferred from {\em Spitzer}.

The variability fraction of 17\,per cent among candidate disks in Sample~II likely reflects a combination of effects. First, Sample~II stars are on average approximately 0.8\,mag fainter in W1 than the Sample~I targets, resulting in larger photometric uncertainties and thus reduced sensitivity to intrinsic flux changes, while at the same time being more prone to source confusion from background sources of the same brightness.  Second, {\em Spitzer} follow-up of a sample of {\em WISE}-selected disk candidates has shown that up to a third are the results of flux contamination, where an unresolved background source -- typically extragalactic -- falls within $1.3\times$ the W1 PSF full width at half maximum, and mimics an infrared excess \citep{Dennihy_2020a}. Such contaminants are generally photometrically stable, and would thereby dilute the measured variability fraction, but could result in a time-dependent component depending on the magnitude and direction of relative proper motion between overlapping sources.

The extent of photometric contamination can be estimated by modelling Sample~II as a mixture of variable and stable white dwarfs, assuming Samples~I and III represent these two populations, respectively. For a given mixture fraction $f_{\rm m}$, a synthetic sample is constructed by randomly drawing a fraction of objects from Sample~I and $1-f$ from Sample~III. The resulting CDF is compared to that of Sample~II using the Kolmogorov-Smirnov statistic, and the fraction that minimizes this metric is adopted, with uncertainties estimated via bootstrap resampling with 2000 iterations. This procedure yields a best-fitting fraction of $f_{\rm m} = 0.49^{+0.17}_{-0.14}$, which at face value suggests around half of the sources are not variable (and thus not dusty).  Assuming this group arises from spurious excesses resulting from source confusion, the implied photometric contamination fraction would be between 0.35 and 0.66, consistent with previous work \citep{Dennihy_2020a}. The properties of Sample~II stars are examined further in the context of source confusion in Section~\ref{sec: candidate analyses}.

\subsection{Characteristics of disk variability}
\label{sec: sample I trends}

\begin{figure}
\includegraphics[width=\columnwidth]{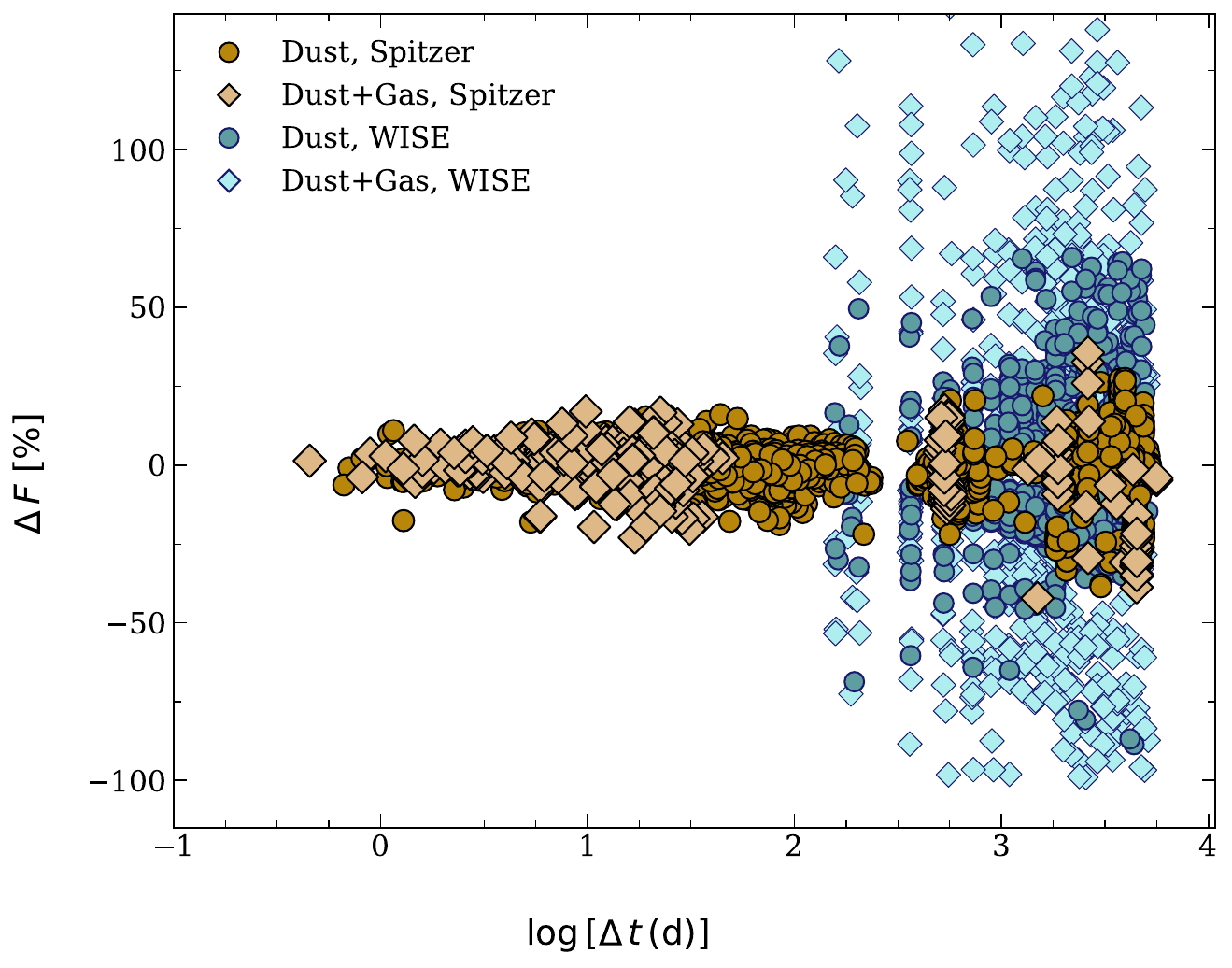}
\caption{Pairwise changes in W1 flux shown in blue for Sample~I targets, and for those disks observed using {\em Spitzer} 3.6\,$\upmu$m in brown.  Debris disks with detected gas emission are shown as diamonds and those without as circles. The distributions of flux changes across both instruments reflect significant differences in cadence, sensitivity, and sample properties owing to selection bias (see text), but are ultimately consistent.  Notably, 0145+234 is not plotted as it would more than double the $y$-axis range and obscure visualization of all other data.}
\label{fig:dfluxvdt}
\end{figure}

Figure~\ref{fig:light curves} plots all W1 light curves for the stars in Sample~I.  While only around one third of these vary above the statistical $5\upsigma$ threshold established in the previous section, it is clear from visual inspection that the most common characteristic is varying degrees of stochasticity.

Figure~\ref{fig:dfluxvdt} presents the distribution of pairwise flux changes in W1 as a function of time baseline for the disks in Sample~I, together with 3.6\,$\upmu$m disk brightness measurements from {\em Spitzer}. Although these datasets probe overlapping wavelength ranges, their distributions differ markedly owing to sensitivity, target selection, and completeness.  {\em Spitzer} detects flux changes as small as a few per cent, and its targeted observations probe a wide range of baselines from minutes to nearly 15\,yr, albeit with irregular cadences and relatively few stars at the longest baselines. In contrast, {\em WISE} provides roughly uniform inter-epoch sampling every 0.5\,yr over a 14.5\,yr baseline, but with lower sensitivity, and thus larger total errors in the epoch fluxes that cannot be circumvented using differential photometry with field sources. As a result, variability on timescales shorter than 0.5\,yr is not sampled, and the observed flux change distribution has a higher proportion of larger amplitudes. 

It has previously been established that changes in dust emission are most pronounced in disks with detectable gas emission, consistent with expectations based on collisions between disk constituents (\citealt{Swan_2020}, \citealt{Guidry_2024}, \citetalias{Noor_2025}).  This phenomenon is also observed here in Sample~I, where these particular systems dominate the extrema in Figure~\ref{fig:dfluxvdt}, which are a factor of 2--$3\times$ larger in amplitude than those observed by {\em Spitzer}. Notably, however, many of these gas-rich debris disks were discovered on the basis of their relatively bright W1 fluxes \citep{Melis_2020,Dennihy_2020b,GentileFusillo_2021b}, which introduces a significant bias into Sample~I as compared to dusty white dwarfs discovered with {\em Spitzer}, which were typically observed on the basis of their metal pollution alone.

Taking this bias into account and excluding disks with detected gas emission, a population-level comparison shows that the peak amplitudes measured with {\em WISE} are broadly consistent with those measured with {\em Spitzer}, and typically on the order of several tens of per cent. Nevertheless, there is a modestly stronger tendency for higher flux changes in {\em WISE}, which are likely due to two factors.  First, many of these dusty and polluted white dwarfs were identified through follow-up spectroscopy of targets with large W1 flux excesses, including many that do not have gas emission \citep{Hoard_2013,Dennihy_2017,Xu_2020}.  This fact favours intrinsically brighter and dustier sources which tend to exhibit higher flux changes, as observed by {\em Spitzer} \citepalias{Noor_2025}.  Second, the roughly uniform observing cadence of {\em WISE} makes it more likely to observe outliers in a distribution as compared to the limited sampling with {\em Spitzer}.

\begin{table}
\centering
\caption{Number of increases and decreases greater than 3$\upsigma$ among the pairwise flux changes measured for infrared time-series samples of disks.  The final column is the relative excess fraction $f_{\rm ex}=(N_+ - N_-)/[(N_+ + N_-)/2]$.}
\label{tab:excess_frac}
\begin{tabular}{lllr} 

\hline

Sample              &$N_{-}$    &$N_{+}$    &$f_{\rm ex}$\\

\hline

\multicolumn{4}{l}{{\em WISE}:} \\     

Sample~I			&786		&537		&$-$0.38\\
Sample~II		    &155		&253		&+0.48\\
Sample~III		    &35\,464	&33\,714	&$-$0.05\\

\\
\multicolumn{4}{l}{{\em Spitzer}:} \\

All			        &325		&294		&$-$0.10\\    
$\Delta t<0.5$\,yr	&207		&213		&+0.03\\   
$\Delta t>0.5$\,yr	&118		&81		    &$-$0.37\\    

\hline

\end{tabular}

{\em Note}.  0145+234 is excluded from these metrics (see Section~3.2).

\end{table}

\begin{figure}
\includegraphics[width=\columnwidth]{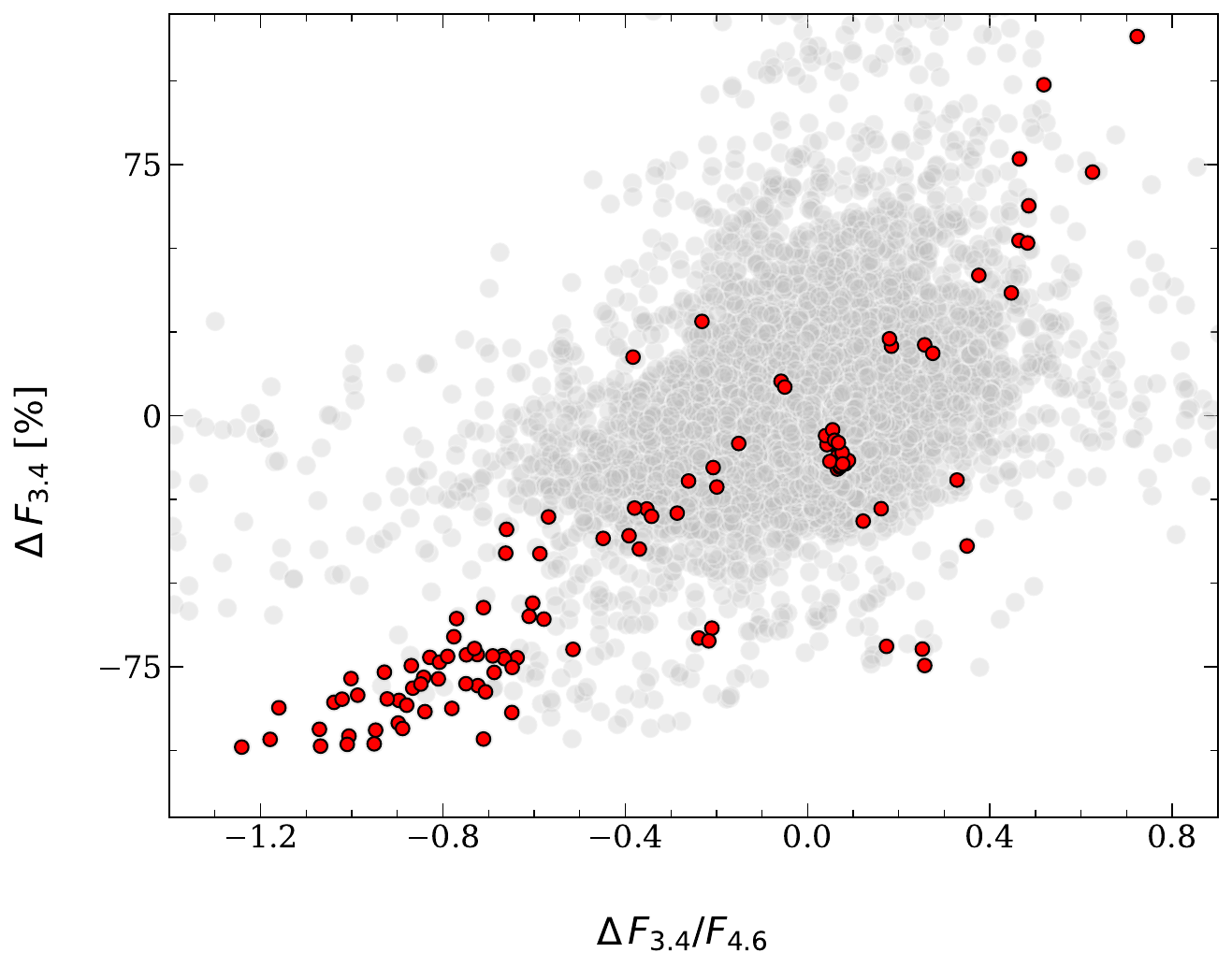}
\caption{Pairwise changes in W1 disk flux vs.\ colour for all Sample~I stars. Grey circles represent individual measurements, and red circles indicate colour changes that are more than $3\upsigma$ significant. Among the significant colour changes, most systems appear redder when dimmer and bluer when brighter, consistent with a change in debris size distribution following planetesimal collisions, where smaller grains that can become hotter than a blackbody are over-represented.}
\label{fig:colvsflux}
\end{figure}

A notable result from these data is that the distribution of flux changes in Sample~I is not symmetric.  There is a clear skew toward flux decreases in both gas-bearing and non-gas systems, where overall there is a 38\,per cent excess of flux decreases, compared with only a 5\,per cent excess for the control stars in Sample~III.  In such an analysis, it is important to exclude 0145+234 as it was discovered via an extreme brightening event in {\em WISE} \citep{Wang_2019} and would thus bias the resulting distribution.  At first glance, this appears to conflict with the symmetric distribution of flux changes reported from {\em Spitzer} disk observations (\citealt{Swan_2020}; \citetalias{Noor_2025}). A closer look reveals that this superficial discrepancy arises from the distinct timescales probed by the two surveys: when the {\em Spitzer} dataset is restricted to the $\Delta t > 0.5$\,yr baselines sampled by {\em WISE}, these data show a (nearly identical) 37\,per cent excess of flux decreases.  On shorter baselines and overall, {\em Spitzer} does indeed exhibit a nearly symmetric or slightly offset distribution of pairwise flux changes (see Table~\ref{tab:excess_frac} for a summary). 

To assess the impact of individual systems on the excess of flux decreases, a leave-one-out test is performed in which each object is removed in turn and $f_{\rm ex}$ recomputed. For Sample~I, $f_{\rm ex}$ remains negative in every realisation, spanning $-0.53 \leq f_{\rm ex} \leq -0.22$, compared with $-0.37$ for the full sample, demonstrating that the asymmetry is not driven by any single system. Together with results from {\em Spitzer}, this behaviour is consistent with gradual dust depletion over approximately year and longer timescales.\\

Despite this tendency toward disk dimming, the infrared excesses in all but two systems persist utterly across the full 14.5\,yr baseline probed by {\em WISE}. In the two remaining systems (0716+404 and 2133+242), the excess is intermittently undetected but subsequently reappears, suggesting the circumstellar dust temporarily falls below the excess detection threshold rather than being fully depleted.

To estimate the physical scale of the observed variability, flux changes are interpreted as variations in the emitting area of the dust disk. Assuming optically thin dust with a power-law, grain size distribution of the form $n(a) \propto a^{-3.5}$ over $0.1\,\upmu$m to 1\,mm, the median and maximum flux changes, and thus dust cross-sections, observed across Sample~I correspond to characteristic changes in dust mass on the order of $10^{18}$ and $10^{21}$\,g, respectively. If attributed to the breakup of a single body with density $\rho=3$\,g\,cm$^{-3}$, these correspond to planetesimals approximately 4 and 50\,km in size. In practice, the observed changes more likely reflect the cumulative effect of multiple events, the detailed modelling of which is beyond the scope of this work.

Figure~\ref{fig:colvsflux} plots pairwise changes in the W1 flux against corresponding colour changes for all Sample~I disks with at least two epochs satisfying the sampling criteria in both {\em WISE} bands. Significant colour changes above $3\upsigma$ are relatively rare among individual targets, as expected given the relatively low S/N of the {\em WISE} observations. Nevertheless, when considering the full ensemble of colour changes, a weak correlation is apparent (Pearson $r=0.20$), where disks tend to appear redder when dimmer and bluer when brighter; this correlation strengthens to $r=0.38$ when considering only those colour changes above $3\upsigma$.  This behaviour corroborates the unambiguous trend established from {\em Spitzer} observations of disks, where $r=0.93$ for $3\upsigma$ or greater colour changes, and is consistent with alterations of the particle size distribution following collisions.

\begin{figure}
\centering
\includegraphics[width=\columnwidth]{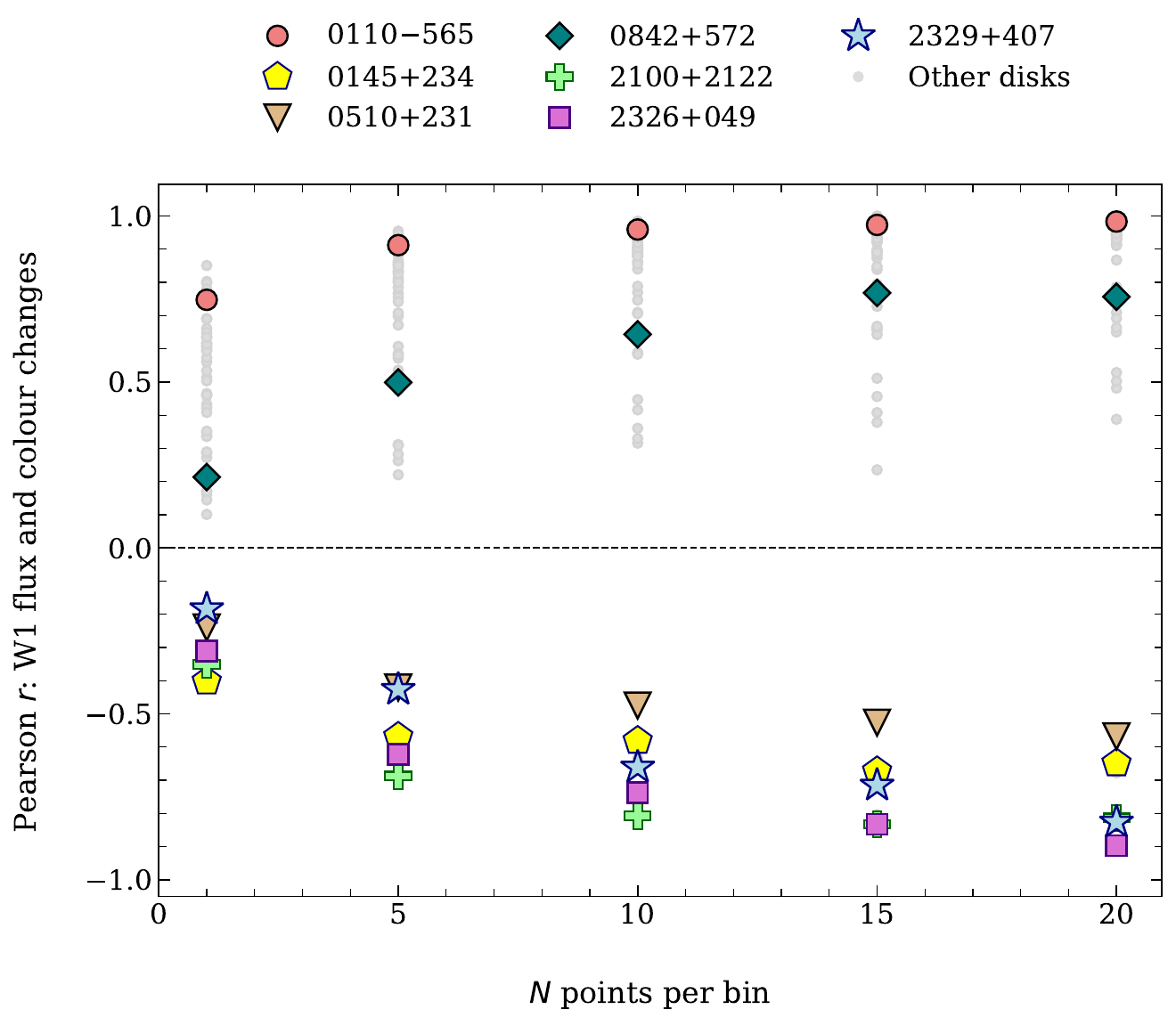}
\caption{Pearson $r$ coefficient between pairwise W1 flux and colour changes as a function of the number of points per bin, where pairwise measurements are sorted by colour change and grouped into consecutive bins of $N$ points before $r$ is computed. $N=1$ corresponds to the unbinned case; for $N>1$, any points remaining after dividing the sorted sample into bins of size $N$ are retained in a final smaller bin. Unique symbols highlight seven systems discussed individually in Sections~\ref{sec: oscillations} and \ref{sec: outbursts}; of these, five show a flux-colour trend that is the inverse of the broader population, becoming redder when brighter and bluer when dimmer, while two follow the typical trend. The remaining 45 Sample~I disks are shown in grey for comparison.}
\label{fig:pearsonr_negatives}
\end{figure}

Notably, when pairwise changes in W1 flux and colour are examined for individual systems, five disks exhibit the inverse correlation, becoming redder when brighter and bluer when dimmer. These are the disks orbiting 0145+234, 0510+231, J2100+2122, 2326+049, and 2329+407. Figure~\ref{fig:pearsonr_negatives} shows the Pearson $r$ coefficient between flux and colour changes for these targets as a function of the number of points per colour change bin, where $n=1$ corresponds to the unbinned case. The inverse flux-colour trend persists across all five systems, irrespective of the binning factor, indicating that it is an intrinsic property of these disks. A closer inspection of {\em Spitzer} data finds that both 0145+234 and 2329+407 exhibit the same inverse trend across baselines ranging from days to several years, but the sparse and irregular sampling of those data preclude a robust comparison with {\em WISE}.

\subsection{Disk luminosity correlates with metal accretion rate}

To further characterize the observed variability in Sample~I, a search is carried out for correlations between variability metrics in both flux and colour, a range of stellar parameters, and disk properties.  Given that source confusion can bias the measured fluxes and fitted disk parameters, this analysis is performed both with and without potentially flux-contaminated sources; in the end, no qualitative differences are found between these.  For nearly all parameter combinations, no statistically significant correlations are found (Pearson $|r| < 0.1$). 

\begin{figure}
\includegraphics[width=\columnwidth]{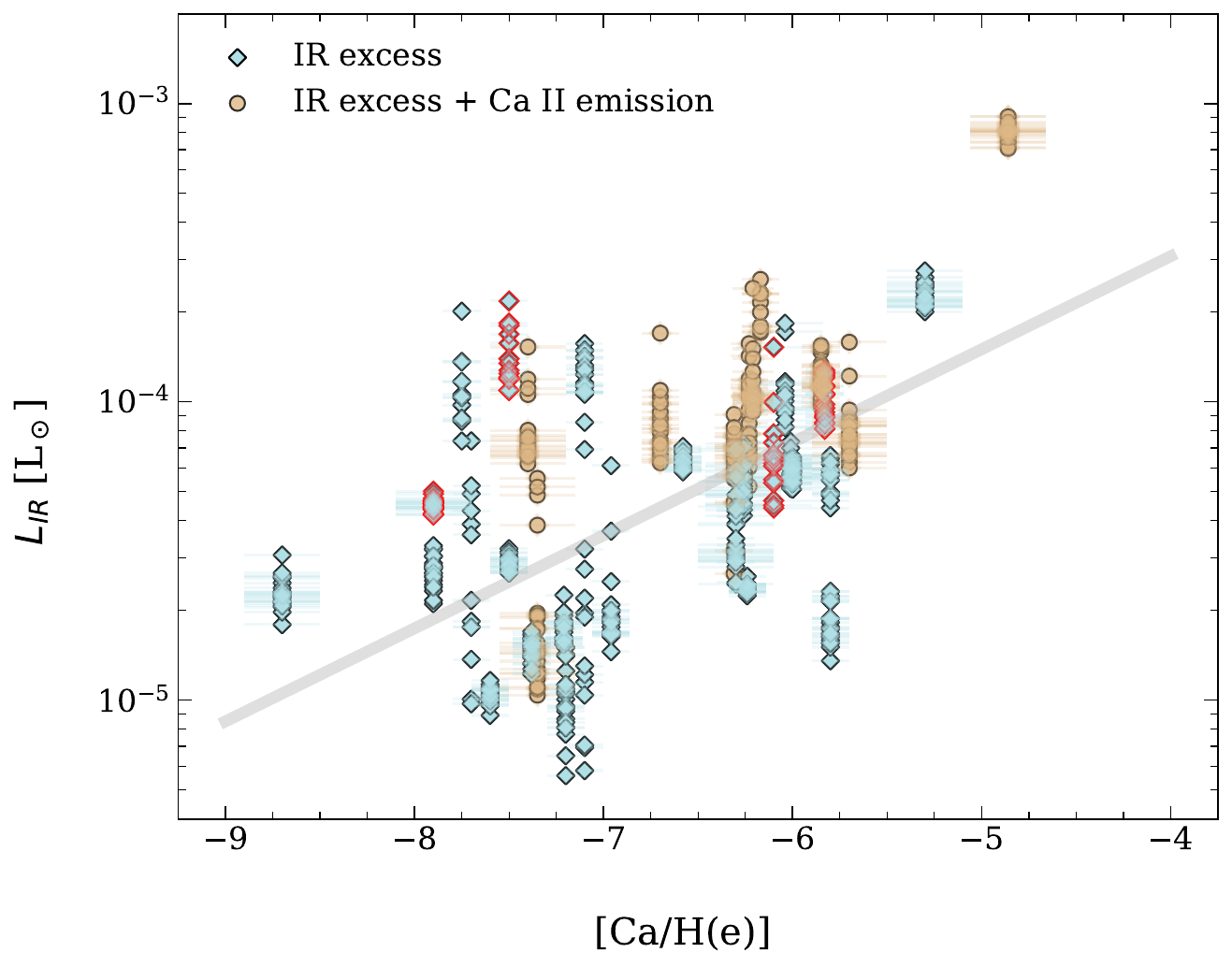}
\caption{Epoch-wise fitted dust disk luminosity against photospheric Ca abundance for all targets in Sample~I. Systems with both infrared excess and gas emission are shown as yellow circles, while those with only infrared excess are plotted as blue diamonds. Sources with potential photometric contamination in the {\em WISE} data are indicated with red outlines. The grey dashed line shows a linear regression fit to the 42 objects that are free of neighbouring sources.}
\label{fig:CavLIR}
\end{figure}

The single exception is illustrated in Figure~\ref{fig:CavLIR}, where a moderately positive correlation is found between fitted disk luminosity and photospheric Ca abundance, with $r = 0.61$.  A nearly identical correlation coefficient and fitted linear slope is found when plotting {\em Spitzer} legacy observations of all white dwarf debris disks using the same parameters \citepalias{Noor_2025}.  The observed correlation is thus real, and underscored by the absence of systems occupying the upper-left region of the figure. Such systems would host the brightest disks in the sample and be readily detectable by any past or present infrared survey, yet none are present despite Sample~I containing disks down to $L_{\rm IR} \sim 10^{-6}\,L_\odot$. The lack of objects in the lower-right region is a result of the fact that there are only a handful of polluted white dwarfs with [Ca/H(e)] $>-6$, some of which are below the detection threshold of {\em WISE}, and where the only two above $-5.5$ are plotted \citep{Williams_2024}. 

This correlation can be interpreted in terms of disk structure and evolution.  In disks that are fully optically thin, the inward drift of particles would nominally be dictated by Poynting-Robertson (PR) drag, and thus it might be expected that white dwarf metal accretion rates are strongly correlated with the total dust masses and hence disk luminosities.  However, even at modest optical depths in white dwarf disks, the inter-particle collision timescale is typically over an order of magnitude shorter than PR drag \citep{Farihi_2008, Farihi_2018}.  Both PR drag and viscous spreading in a particulate disk should increase with surface density when collisions are not damped \citep{Bromley_2015}, and Figure~\ref{fig:CavLIR} is consistent with these expectations without distinguishing between the two mechanisms.  Once all solids are sufficiently close to the star to be sublimated, a fully gaseous disk will be subject to the bottleneck of viscous spreading prior to accretion \citep{Metzger_2012, Farihi_2012}.  Depending on this timescale, the mass accretion rate onto the white dwarfs can be either higher or lower than that dictated by PR drag, depending on the history upstream.  Nevertheless, the correlation indicates that, on average, the inward drift of particles increases with increasing dust mass.

\subsection{Effects of source confusion}
\label{sec: candidate analyses}

In addition to the substantially weaker evidence for variability, the disk candidates in Sample~II display markedly different overall properties to the confirmed disks studied here in Sample~I and those observed by {\em Spitzer}.  In particular, there is ample evidence for source confusion.

First, the pairwise flux changes are drastically skewed toward increases, where Table~\ref{tab:excess_frac} shows their collective behaviour is diametrically opposed to the confirmed disk samples of {\em Spitzer} and {\em WISE}.  A leave-one-out test confirms that this positive asymmetry is not driven by any single object and is an intrinsic property of the sample, with $f_{\rm ex}$ remaining positive in every realisation. This alone suggests Sample~II may be dominated by spurious detections of infrared excess, or where a genuine excess is not caused by circumstellar dust.  Second, the objects in Sample~II have an elevated source density, where around 90\,per cent have a neighbouring source within 7.8\,arcsec (cf.\ 20\,per cent of Sample~I stars), suggesting that background galaxies contribute to their W1 light curves and may produce spurious, time-dependent infrared excesses whose light curves are distinct from those manifested by genuine debris disks. The positive $f_{\rm ex}$ excess might conceivably result from the relative proper motion between a white dwarf and neighbouring source, where photometric contamination increases over time as the separation narrows over the {\em WISE} mission baseline.  However, the exact mechanism responsible for the positive asymmetry is beyond the scope of the present work.

The probability that a given candidate excess originates from a background galaxy can be estimated using source counts from {\em Spitzer} IRAC observations at 3.6\,$\upmu$m \citep{Fazio_2004b}, as these should be similar in W1 at 3.4\,$\upmu$m. For the median W1 excess magnitude of 16.6\,mag in Sample~II, the corresponding IRAC galaxy counts are $10^{3.9}$ per magnitude per square degree.  Adopting a 7.8\,arcsec radius, the probability of a sufficiently bright background galaxy falling within this area is 4.1\,per cent, increasing to 11\,per cent for the faintest excesses in Sample~II.  Because these candidates are selected from initial samples of 6000 to 8500 {\em Gaia} white dwarfs \citep{Xu_2020,RebassaMansergas_2019}, background galaxies are sufficient to account for all of Sample~II on the basis of statistics alone.  

Interestingly, the Sample~I disks with neighbouring sources also display a strong skew toward flux decreases, with $f_{\rm ex}=-0.78$.  This suggests that, by itself, having a nearby source on the sky is insufficient to change the character of long-term, genuine disk variability.  
However, when the candidate infrared excess sources are themselves a product of crowding, the collective properties of the sample differ from those of genuine disks.
Altogether, the analysis here indicates that Sample~II stars are, on average, unlikely to host genuine dusty disks. However, there may be some genuine disks among the candidates, but these require infrared data at higher spatial resolution.

\subsection{Oscillation in 2326+049 and other disks}
\label{sec: oscillations}

\begin{figure}
\includegraphics[width=\columnwidth]{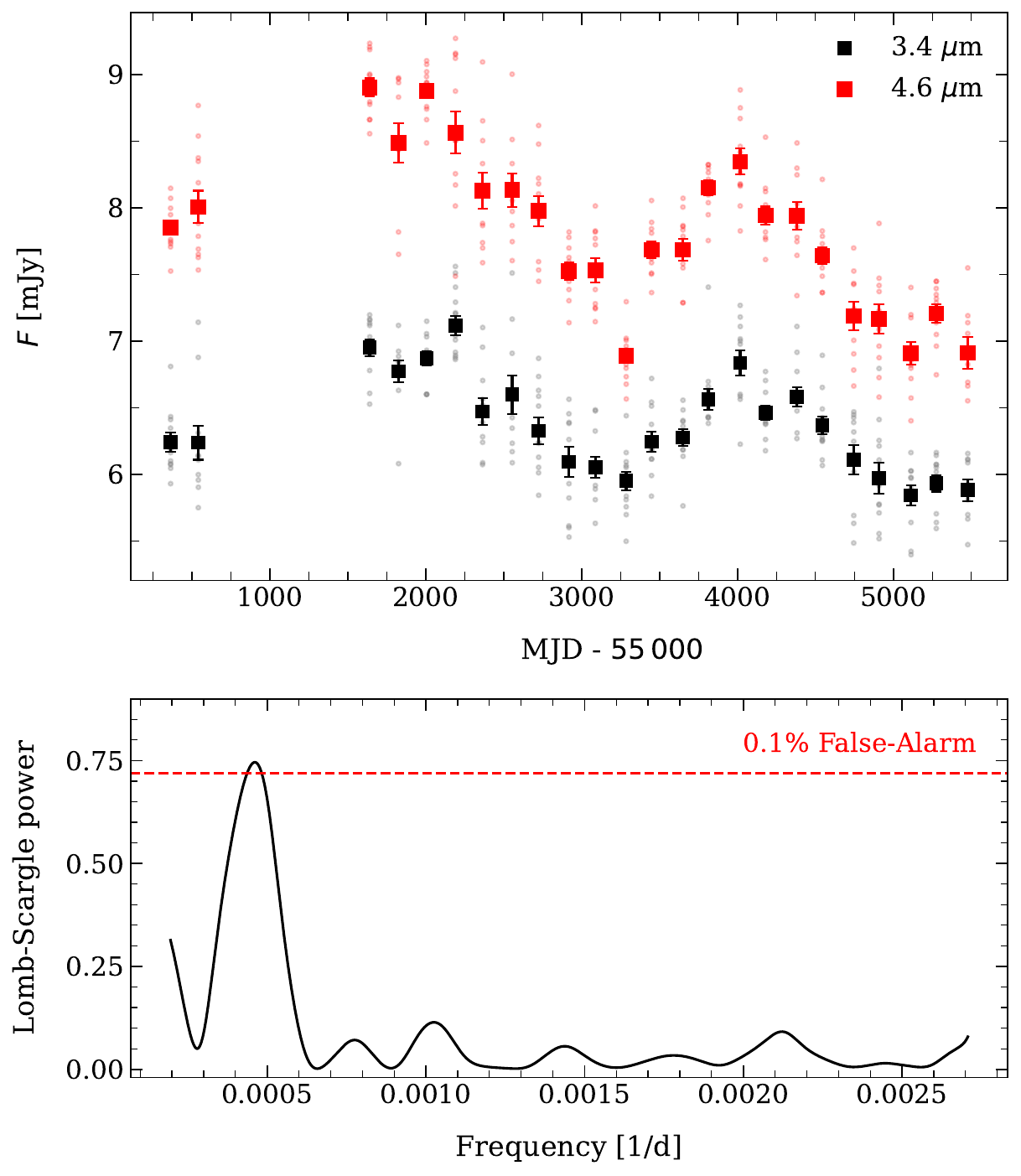}
\caption{The top panel shows the W1 (black) and W2 (red) light curves for the dust disk orbiting 2326+049.  Smaller and lighter-coloured circles show individual flux measurements, while darker squares indicate the weighted, epoch-mean fluxes and associated uncertainties.  In the lower panel is the corresponding Lomb-Scargle periodogram of the W1 light curve, where the peak frequency 0.00046\,d$^{-1}$ corresponds to a period of 5.9\,yr, and where the power exceeds a bootstrap computed 0.1\,per cent false alarm, where this threshold is marked by a red dashed line.}
\label{fig:LSP}
\end{figure}

Five disks have light curves with varying degrees of oscillation-like morphology, from relatively convincing to speculative, each identified by visual inspection; these are 0110$-$565, 0510+231, 0842+572, 2326+049, and 2329+407.  All five have strong inter-band agreement between W1 and W2 (Pearson $r > 0.9$).  Of these, 2326+049 (the prototype dusty white dwarf G29-38) stands out as the only target with an outstanding signal peak in a Lomb–Scargle periodogram of its W1 light curve, as shown in Figure~\ref{fig:LSP}.  The peak frequency corresponds to a 5.9\,yr period and is formally significant above 99.9\,per cent confidence. However, the light curve samples only two cycles, with a substantial gap between the cryogenic and re-activated {\em WISE} missions, and thus the data cannot establish any signal longevity. Although this star is a known ZZ Ceti pulsator, its minute-timescale and few-per cent brightening episodes \citep{ShulovKopatskaya_1974,McGraw1975,Kleinman1998} should leave no lasting footprint in the day-long averaged {\em WISE} visit. Although pulsation-driven, disk-brightness variations in the near-infrared are well-established \citep{Graham_1990, Reach_2009, vonHippel_2024}, any changes on these minute timescales cannot account for the long-term variations observed in {\em WISE}.

An attempt is made to characterize oscillation timescales in the other four sources by fitting sine waves to their W1 light curves.  However, these fits fail to capture the structure of the time-series data, indicating that the variability clearly departs from periodic.  Possible characteristic timescales are instead estimated from the mean interval between successive flux minima, identified using the \textsc{scipy.signal.find\_peaks} package, and applied to W1 light curves smoothed with a one-dimensional Gaussian filter with kernel width 0.5\,yr to reduce short-term fluctuations between neighbouring epochs.  Across all five systems, the crudely estimated timescales fall in the range 3--6\,yr.  If this is interpreted as Keplerian orbital periods, these correspond to semimajor axes of 2--3\,au.  

The flux-colour trends offer a further means of characterizing this subset. Two systems, 0110$-$565 and 0842+572, display a positive correlation in line with the population-wide trend, whereas the remaining three, 0510+231, 2326+049, and 2329+407, instead become redder when brighter and bluer when dimmer, placing them among the five disks that exhibit this inverse correlation (Figure~\ref{fig:pearsonr_negatives}). The physical origin of this behaviour is considered in Section~\ref{sec:qp explanation}.

\subsection{A new outburst and the post-outburst flux of 0145+234}
\label{sec: outbursts}

\begin{figure}
\includegraphics[width=\columnwidth]{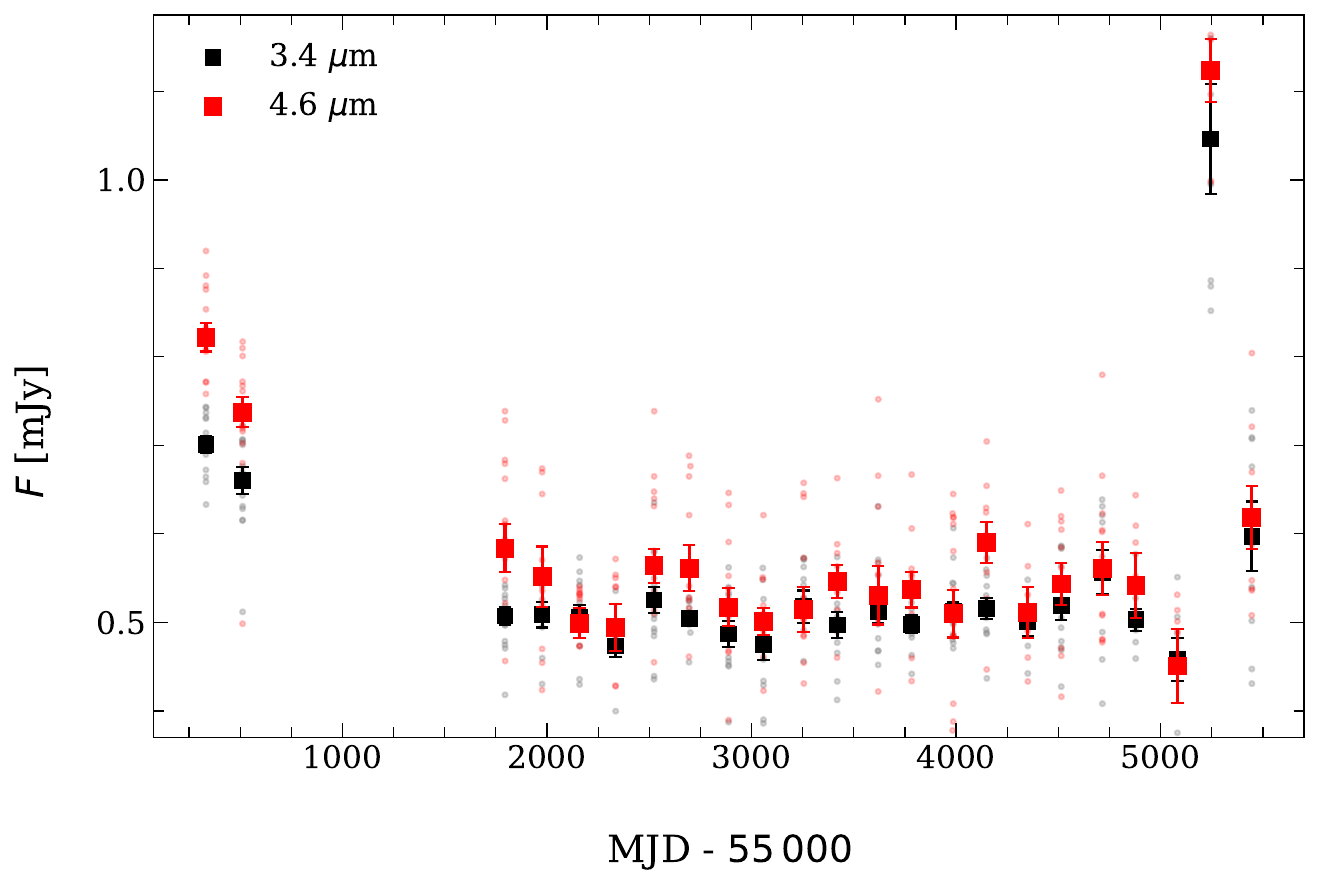}
\vskip 3 pt
\includegraphics[width=\columnwidth]{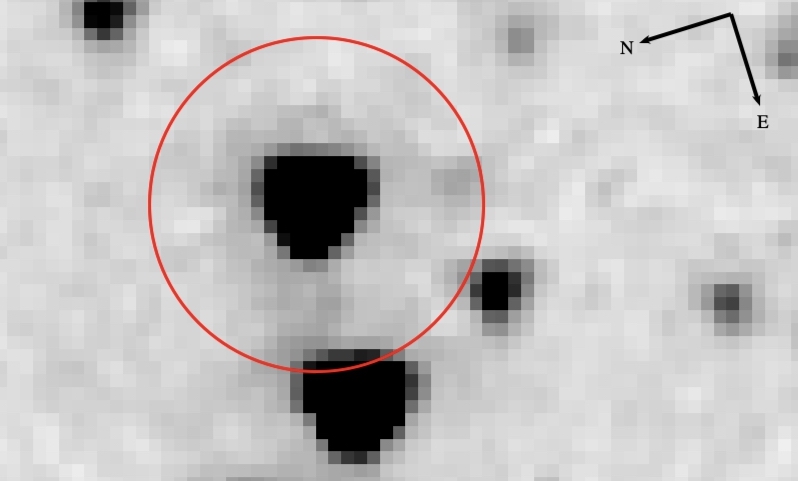}
\caption{The upper panel plots W1 and W2 light curves for the dust disk at J2100+2122, with symbols identical to Figure~\ref{fig:LSP}.  A candidate outburst is visible near MJD 60\,250 (2023 November), during which the disk flux more than doubled in both bands over 160\,d.  The lower panel is a {\em Spitzer} image of J2100+2122 obtained in 2019 September, where the red circle has a radius of 7.8\,arcsec = $1.3\times$ the W1 point-spread function full width at half maximum.  The comparably bright source just outside this radius has an independent photometric fit and is thus actively de-blended in all datasets used for the light curve (Section~\ref{sec:sample_selection}).  The mutual proper motion of the two stars has kept the pair separated by more than 9.1\,arcsec for all {\em WISE} epochs.}
\label{fig:wdj2100}
\end{figure}

A single-epoch outburst is identified in the light curves of J2100+2122, where the W1 and W2 disk fluxes increase by factors of 2.3 and 2.6, respectively, as shown in the upper panel of Figure~\ref{fig:wdj2100}.  The brightening takes place within 160\,d and is comparable in timescale and amplitude to the brightening observed at 0145+234, where the disk-only fluxes in both bands increased by a factor of four over 200\,d \citep{Wang_2019}. Following the sudden rise, the disk flux of J2100+2122 declines substantially by the subsequent and final {\em WISE} epoch, where it remains modestly above the pre-outburst baseline. Although based on a single post-outburst epoch, this resembles the post-outburst behaviour of 0145+234 (Figure~\ref{fig:light curves}) and may suggest that the disks have settled into a new equilibrium with a greater surface area of micron-sized dust particles. While the single-epoch, post-outburst sampling of J2100+2122 prevents a determination of a decay timescale, the decline visually appears more rapid than that of 0145+234, with any physical difference potentially reflecting variations in the properties of the debris produced by individual events, including the particle size distribution and mass density \citep{KenyonBromley_2017a}.

As a measure of caution, given a single flare in an otherwise relatively featureless time-series several years prior, the possibility of source confusion is assessed for J2100+2122. In fact, there is a comparably bright background star present at a separation of 9.6\,arcsec in a {\em Spitzer} IRAC image of this source, as shown in the lower panel of Figure~\ref{fig:wdj2100}.

However, all indications support an intrinsic brightening for the disk orbiting J2100+2122. First, the relative proper motion has gradually brought the pair closer, with separations greater than 10.4\,arcsec during the first epoch in Figure~\ref{fig:wdj2100}, 9.3\,arcsec at the outburst epoch and 9.1\,arcsec in the subsequent and final {\em WISE} epoch.  Consistent source blending is expected to yield a monotonic trend with decreasing separation but this is not observed.  Second, the neighbouring source has a stable {\em WISE} light curve, with no variation coincident with the disk brightening.  Third, in nearly all {\em WISE} detections, the pipeline fitted two PSF components, indicating that the target and neighbour are actively de-blended, with the {\em WISE}-detected positions as expected from {\em Gaia} astrometry \citep{Debes_2011}.

It is notable that independent, ground-based monitoring of J2100+2122 reveals two $K_s$-band brightening episodes of order 50\,per cent approximately 500 and 100\,d prior to the outburst observed by {\em WISE}, both of which decay over timescales of order weeks \citep{Rogers_2025a}. The event identified here may therefore represent only one of several, large-amplitude brightening episodes, with shorter-lived events capable of occurring and fading entirely between the six month {\em WISE} sampling.

With source confusion excluded, the properties of the {\em WISE}-detected outburst at J2100+2122 can be examined further and compared directly with 0145+234. Across the entire light curve, the fitted dust temperature for J2100+2122 remains constant within the uncertainties, with a mean of 1000\,K and a $1\upsigma$ scatter of 70\,K, while the inferred emitting surface area increases by a factor of approximately four during the flare. For comparison, the emitting area of dust toward 0145+234 increases by a factor of 18 between its low and high states. Assuming a power-law grain size distribution of the form $n(a)\propto a^{-3.5}$ between 0.1\,$\upmu$m and 1\,mm, and adopting a grain density of $\uprho=3$\,g\,cm$^{-3}$, the dust masses required to account for the absolute increases in emitting area are approximately $10^{19}$\,g for J2100+2122, compared to $10^{18}$\,g for 0145+234. These estimates fall at the upper end of the dust mass changes inferred from flux variations across Sample~I, but remain within the range spanned by the broader population (Section~\ref{sec: sample I trends}). 

Notably, both 0145+234 and J2100+2122 host detectable gas disks and exhibit the largest fractional flux variations in Sample~I, consistent with gas-bearing systems representing a more dynamically active subclass of white dwarf debris disks \citep{Swan_2020}. Both sources also display the inverse flux-colour trend to the broader population of disks in Sample~I, becoming redder when brighter and bluer when dimmer (Figure~\ref{fig:pearsonr_negatives}), and this correlation persists regardless of whether the outburst epochs are included in the analysis.

\subsection{Long-term decay in \texorpdfstring{0408$-$041}{0408-041}}
\label{sec: decaying disks}

The light curve of 0408$-$041 (Figure~\ref{fig:light curves}) displays a clear trend of decaying flux from an unspecified maximum that likely occurred between the cryogenic and re-activated {\em WISE} missions \citep{Farihi_2018}. The infrared fluxes in both {\em WISE} bands closely correlate, and the decline appears to become horizontal a few epochs prior to the end of mission.  The normalized light curves are fitted with a function $F = F_0/(1+t/t_{\rm d})$, where $F$ is the disk flux at time $t$, $F_0$ is the initial flux at the first re-activation epoch, and $t_{\rm d}$ is the decay timescale. Each {\em WISE} band is well described by a decay timescale of approximately 700\,d.  
Similar behaviour is also seen in the post-outburst light curve of 0145+234, as observed by {\em Spitzer} before the end of its mission and later with {\em JWST}/MIRI \citep{Swan_2021, Swan_2024}. Fitting those data with the same functional form yields a decay timescale of approximately 180\,d in both {\em Spitzer} bandpasses at 3.6 and 4.5\,$\upmu$m, corresponding to a decline on the order of one year.

There may be a second disk in the full {\em WISE} dataset that exhibits similar and correlated trend of decreasing fluxes. The W1 light curve of 2133+242 (Figure~\ref{fig:light curves}) may show decay that either starts or continues from the two cryogenic mission epochs, and which may persist for the first several epochs after re-activation, then becomes flatter but with no coherent pattern (the W2 data have low S/N to assess).  However, the evidence for this disk is far less clear, and it is not possible to be certain that a decay trend continues through the sizable observing gap.  However, assuming the decreasing trend continues across the dates where there are no data, a decay timescale of approximately 500\,d is derived.

\section{Discussion}
\label{sec:discussion}

The following discussion addresses the physical interpretation of the variability trends established above, first considering the origin of the asymmetric flux distribution, then the flux-colour dichotomy among Sample~I disks, and finally the prospects for future observations.

\subsection{Origin of asymmetric flux changes}
\label{sec: implications}

The baseline-dependent asymmetry toward flux decreases can be accounted for within the framework of collisional cascades. In a collisional cascade, the particle population evolves as $N(t) = N_0/(1+t/t_{\rm col})$ \citep{DominikDecin_2003}, where the collisional timescale $t_{\rm col}$ depends on particle size, with smaller grains evolving more rapidly than larger fragments. For grains dominating some optical depth $\uptau$, this timescale can be approximated as $t_{\rm col} \approx P/(4\pi\tau)$, where $P$ is the orbital period. For disks near the Roche radius ($P\approx5$\,h) and $\uptau\approx10^{-3}$--$10^{-2}$ as inferred from the observed 3--5\,$\upmu$m excesses \citep{Rocchetto_2015}, this yields $t_{\rm col} \sim$1--20\,d for micron-sized grains. At the higher cadence of {\em Spitzer}, which probes timescales of minutes to weeks, $\Delta t \lesssim t_{\rm col}$: the collisional evolution of individual grain populations is resolved, dust production and destruction events are captured in roughly equal measure, and the flux change distribution is consequently symmetric. At baselines greater than 0.5\,yr, in contrast, $\Delta t \gg t_{\rm col}$ for micron-sized grains, and their rapid evolution is effectively averaged out. The observed variability on these longer baselines instead reflects the interplay between intermittent dust injection and continuous collisional decay. When mass input is infrequent relative to the decay rate, the system spends the majority of its time in a declining state, and flux decreases are statistically more likely to be observed. This picture is consistent with numerical simulations of collisionally evolving disks, both around white dwarfs and younger stars, which predict precisely this combination of short-term stochastic variability and longer-term net fading when mass supply is intermittent or absent \citep[e.g.\ ][]{DominikDecin_2003, Gaspar_2013, Wyatt_2014, KenyonBromley_2017a}.

Within the same framework, the  observed decay timescales  provide a crude constraint on the largest bodies feeding the cascade. The population-wide tendency toward flux decreases on baselines longer than 0.5\,yr, together with the $\sim 1$\,yr declines measured for individual systems (Section~\ref{sec: decaying disks}), is broadly consistent with the decay of kilometre-scale swarms in collisional cascade simulations \citep[Fig.~5]{KenyonBromley_2017a}. This scale is also theoretically motivated, as substantially larger bodies are less likely to survive tidal disruption owing to insufficient material strength, while kilometre-scale fragments are expected to persist and continue to feed the cascade \citep{KenyonBromley_2017a, Brouwers_2021, Steckloff_2026}.

Two additional dust removal processes -- sublimation and PR drag -- are worth considering but are unlikely to be the primary drivers of this asymmetry. Collisional grinding can produce grains sufficiently small to sublimate under irradiation from the white dwarf, thereby providing a possible sink. However, if this were the dominant dust removal mechanism, a correlation between variability amplitude and stellar effective temperature would be expected, and none is observed in Sample~I. PR drag may also contribute to dust removal \citep{Rafikov_2011, BochkarevRafikov_2011}, but for micron-sized grains at the Roche radius, collisional timescales are at least an order of magnitude shorter than the PR drag timescale \citep{Farihi_2008, Farihi_2018}, such that grains are more likely to be destroyed before undergoing significant radial migration. Both processes likely operate at some level, but quantifying their interplay with collisional evolution is beyond the scope of the present work.

\subsection{The flux-colour change dichotomy among disks}
\label{sec:qp explanation}

The flux-colour correlations of the {\em WISE}-observed disks provide key constraints on the origin of the variability. Both disks with outbursts follow a trend opposite to the Sample~I overall trend, and become redder when brighter and bluer when dimmer.  Similarly, of the five systems with possible oscillations, two follow the population-wide trend, while the remaining three display the inverse correlation.  Both groups displaying this behaviour were identified as oscillating or outbursting based solely on visual inspection, but their strong Pearson $r$ correlations in Figure~\ref{fig:pearsonr_negatives} indicate at least some underlying processes in common.  Notably, as in all five of these systems, the inverse colour trend present in the outbursting disks 0145+234 and J2100+2122 {\em persists across all epochs and is not confined to the brightening events}.

The  framework of planetesimal collisions invoked to account for the widespread infrared variability of dusty white dwarfs (\citealt{Farihi_2018, Swan_2020}, \citetalias{Noor_2025}), provides a natural starting point for both groups. When a body on an eccentric orbit impacts a pre-existing disk, or in any orbit-crossing collision, the resulting debris is distributed across a range of orbital elements, but the post-impact trajectories continue to intersect at this fixed collision point \citep{JacksonWyatt_2012, Jackson_2014}. A collisional cascade within this newly generated debris can replenish the micron-sized grain reservoir and produce a corresponding rise in infrared flux \citep[e.g.][]{Wyatt_2008}. The rate at which small grains accumulate depends on the mass density of the debris cloud \citep{Su_2019}, while the subsequent decline in emission reflects the depletion of the largest fragments sustaining the cascade, such that populations dominated by smaller bodies are processed rapidly, whereas larger surviving fragments can feed dust production over longer intervals \citep{Wyatt_2007b, Gaspar_2013, KenyonBromley_2017a}. Successive brightening episodes are therefore not expected to be strictly periodic or uniform in amplitude, as each is governed by the mass, geometry, and particle size distribution of the colliding bodies.

For the systems that follow the population-wide colour trend, this framework provides a straightforward account of both the flux and colour evolution: a collision temporarily skews the size distribution toward smaller grains and produces hotter, bluer emission, followed by a gradual return toward the steady-state size distribution as the cascade evolves \citepalias{Noor_2025}. Repeated collisions would then drive the episodic brightening and fading, including the observed oscillation-like morphology. A related possibility is that these collisions occur preferentially near periastron, where dust is produced at higher temperatures than material generated further from the star. Enhanced production at this location would yield both brighter and bluer emission, with a subsequent decline and reddening as this hotter component is ground down. Orbital motion of inhomogeneously distributed debris after a tidal disruption event could in principle produce similar colour modulation, but the debris stream is expected to fill out azimuthally within tens of orbits \citep{Nixon_2020}, making it unlikely that multiple systems would be caught during this short-lived phase.

In all five disks that become redder when brighter and bluer when dimmer, the flux increase must be accompanied by an increased contribution from dust that is cooler than the pre-existing population. A natural site for the production of such material is near apastron, where newly generated dust lies further from the white dwarf and is therefore cooler. Collisions at this location may be promoted if the disk is precessing, as inferred for disks with detected gas \citep[e.g.][]{Manser_2016, Manser_2021, MirandaRafikov_2018}, in which case its misalignment with incoming debris would facilitate orbit crossings and destructive collisions at larger radii \citep{Brouwers_2021}. This picture can be applied to 2326+049, whose W1 light curve exhibits variability on a timescale of approximately 6\,yr (Figure~\ref{fig:LSP}). The timescale is consistent with an eccentric fragment that intercepts the pre-existing disk near apastron once per orbit. Repeated passages through this region would preferentially generate cooler dust, producing redder and brighter emission, before collisional grinding depletes the newly formed grains and the system fades back toward a bluer state.

Taken together, these scenarios are qualitatively consistent with the observed trends, but detailed modelling is required to determine whether they can reproduce the observed amplitudes, timescales, and colour evolution.

\subsection{The end of an era for infrared time-domain studies}

The monitoring enabled by {\em WISE} has been invaluable for the study of white dwarf debris disks, providing nearly 15\,yr of relatively uniform mid-infrared cadence, and establishing variability as a widespread property of these systems \citep{Swan_2019}. Together with {\em Spitzer}, these missions define an era of time-domain disk monitoring for white dwarfs, delivering an invaluable legacy of multi-epoch observations at 3--5\,$\upmu$m that is best suited for detecting warm dust near the stellar Roche radius. With the conclusion of {\em WISE}, that era has effectively ended, and further progress in characterizing debris disk evolution must now rely on the next generation of observing facilities.

{\em JWST} has already begun delivering results for polluted and dusty white dwarfs \citep{Swan_2024,Farihi_2025}, offering unprecedented sensitivity and spectral resolution. However, the substantial pointing overhead per target renders it less suited to population-wide, time-domain monitoring previously enabled by either {\em Spitzer} or {\em WISE}. {\em SPHEREx} provides all-sky spectrophotometric coverage from 0.75 to 5\,$\upmu$m, with a cadence of approximately 0.5\,yr \citep{Crill_2020}, thus duplicating the cadence and extending the wavelength coverage of {\em WISE}.  However, observations with the six linear variable filters are not simultaneous, and typically require 2\,weeks for complete spectral coverage \citep{Bock_2026}, thus complicating any interpretation.  The effective PSF is similar in size to that of {\em WISE}, limiting its utility in crowded fields, and early indications suggest that it will primarily be sensitive to the brightest subset of white dwarf debris disks.

Roman should provide weekly monitoring of an 18\,deg$^2$ field over two years, but only for wavelengths up to 2.3\,$\upmu$m, which may be impacted by the thermal emission of the telescope \citep{Schlieder_2024}.  Nevertheless, the vast majority of white dwarfs with detected excess at the longest wavelengths will be too faint for {\em Gaia}, and thus challenging to distinguish disks from binaries and neighbouring background sources.  $K$-band light curves for any bright disks in this field will be valuable, but the majority of disks emit weakly or not at all at these wavelengths. The {\em NEO Surveyor} will image in two filters centred near 4 and 8\,$\upmu$m over a planned five year mission \citep{Mainzer_2023}, but limited to the ecliptic plane and an observing window of several days, best suited to detect the motion of nearby objects. 

Infrared variability studies and dynamical considerations have made it clear that new material is periodically injected into white dwarf debris disks, and which likely originates in a cold planetesimal belt.  Although thermal detections of colder debris at large orbital distances have remained elusive \citep{Farihi_2014}, episodic transiting debris clouds are observed in a modest but growing number of systems, with durations and inter-episode timescales that suggest orbits that are orders of magnitude more distant than those typically inferred from the 3--5\,$\upmu$m excesses \citep{Vanderbosch_2020,Guidry_2021,Bhattacharjee_2025}. Continued monitoring of such systems, along with further transit discoveries, may enable constraints on debris in the outer planetary systems of white dwarfs, while targeted searches using {\em JWST} and {\em ALMA}, as well as the far-infrared facility {\em PRIMA} could potentially detect thermal emission from cold debris belts, though success will depend on whether there is sufficient mass and surface density.

\section{Conclusions}
\label{sec:conclusion}

This study analyzes multi-epoch photometry from {\em WISE}, spanning the full 14.5\,yr duration of the mission to investigate variability in emission from confirmed and candidate dusty white dwarfs (Samples~I and~II, respectively). Variability is widespread within Sample~I and, after accounting for the reduced photometric sensitivity of {\em WISE} relative to {\em Spitzer}, is consistent with being ubiquitous. In contrast, Sample~II exhibits markedly different variability behaviour, failing to reproduce key trends observed in Sample~I, indicating that a substantial fraction of these candidates may be spurious, likely as a result of source confusion within the {\em WISE} beam.

Key results for Sample~I are summarized as follows:
\begin{itemize}

\item{{\em WISE} confirms the trend of flux changes that are typically redder when dimmer and bluer when brighter, first identified by {\em Spitzer} time-series observations.}

\item On baselines longer than 0.5\,yr, including all {\em WISE} measurements and a subset of {\em Spitzer} baselines, flux changes are significantly skewed toward decreases, consistent with gradual dust depletion.

\item Disk evolution is stochastic on many baselines, and on multi-year timescales there can be possible oscillations, rapid brightening, and long-term decay (each clearly observed in at least one case).

\item A moderately positive correlation is observed between dust luminosity and photospheric calcium abundance, consistent with the accretion of optically thin disk material.

\item Source confusion is always important when identifying genuine disks using {\em WISE} infrared excesses toward white dwarfs.

\end{itemize}

Together with the shorter-cadence {\em Spitzer} monitoring \citepalias{Noor_2025}, these results demonstrate that variability in white dwarf debris disks is ubiquitous, spanning timescales from minutes to decades, while the infrared excesses themselves persist throughout the 14.5\,yr {\em WISE} baseline. As the only facility to provide roughly uniform mid-infrared monitoring on such timescales, {\em WISE} has given a unique, population-level view of white dwarf debris disk evolution, bridging short-timescale variability with long-term trends. While current and next generation infrared telescopes may offer gains in sensitivity, spectral coverage, and spatial resolution, their ability to provide comparable long-baseline monitoring remains uncertain, underscoring the lasting legacy of {\em WISE} in placing empirical constraints on debris disk evolution around white dwarfs.

\section*{Acknowledgements}

The authors thank the anonymous referee whose comments helped improve the manuscript. H.~T.~Noor thanks M.~C.~Wyatt for helpful discussions on impact-driven variability. This study makes use of data products from the (Near-Earth Object) Wide-field Infrared Survey Explorer, which is a joint project of the University of California, Los Angeles, and the Jet Propulsion Laboratory / California Institute of Technology, funded by the National Aeronautics and Space Administration.

\section*{Data Availability}
The data underlying this study are publicly available through the NASA/IPAC Infrared Science Archive (IRSA) at \url{https://irsa.ipac.caltech.edu}.



\bibliographystyle{mnras}
\bibliography{refs} 


\appendix

\section{{\em WISE} Light curves}
\label{sec: light curves}

The light curves of all well-sampled targets are given below.

\begin{figure*}
\vskip 1cm
\includegraphics[width=\textwidth]{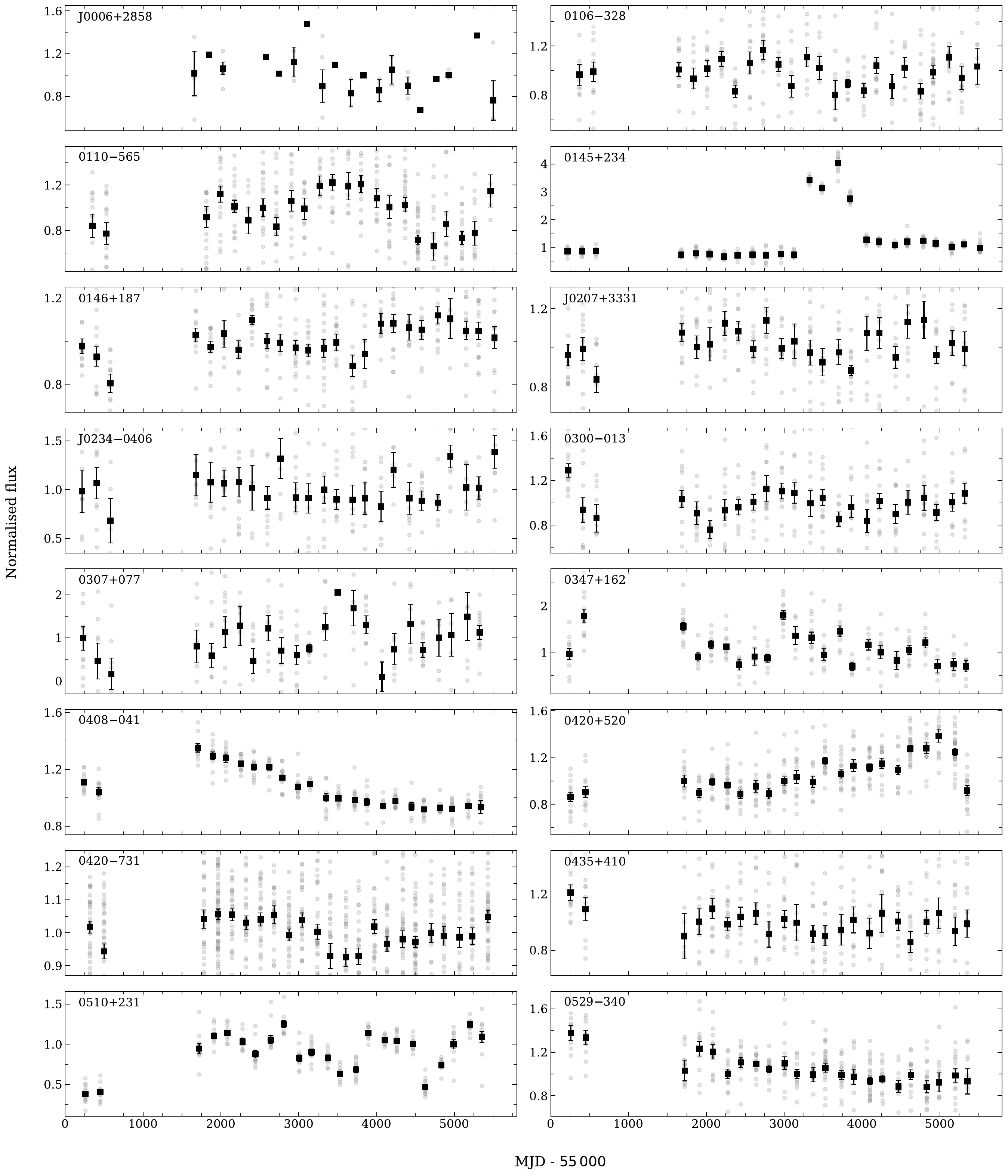}
\caption{Normalized {\em WISE} light curves for Sample~I relative to the median flux. W1 light curves are shown in black and W2 in red; the latter generally have lower S/N. Symbols are the same as in Figure~\ref{fig:LSP}.}
\label{fig:light curves}
\end{figure*}

\addtocounter{figure}{-1}
\begin{figure*}
\vskip .1cm
\includegraphics[width=\textwidth]{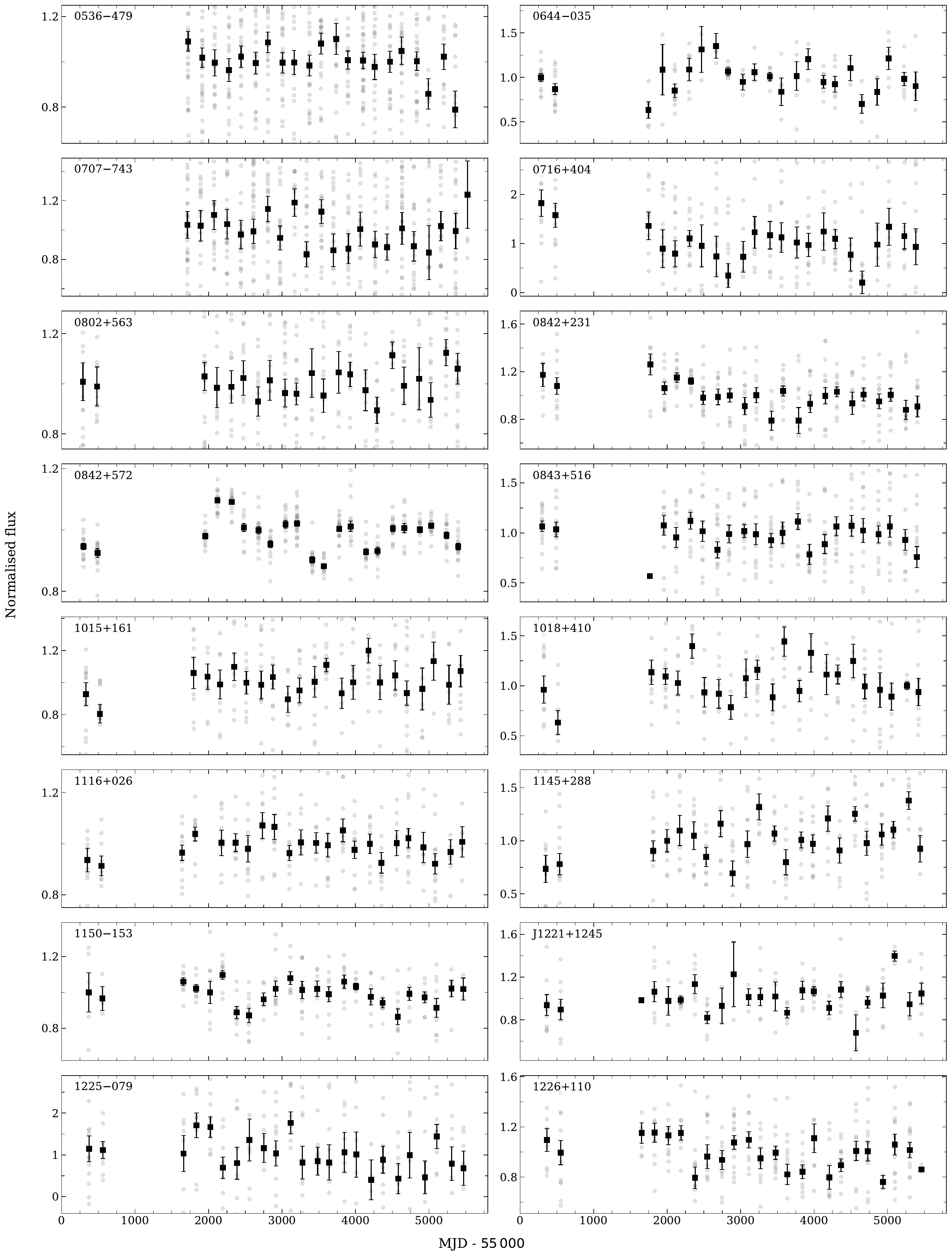}
\vskip -.2cm
\caption{ Cont.}
\end{figure*}

\addtocounter{figure}{-1}
\begin{figure*}
\vskip .1cm
\includegraphics[width=\textwidth]{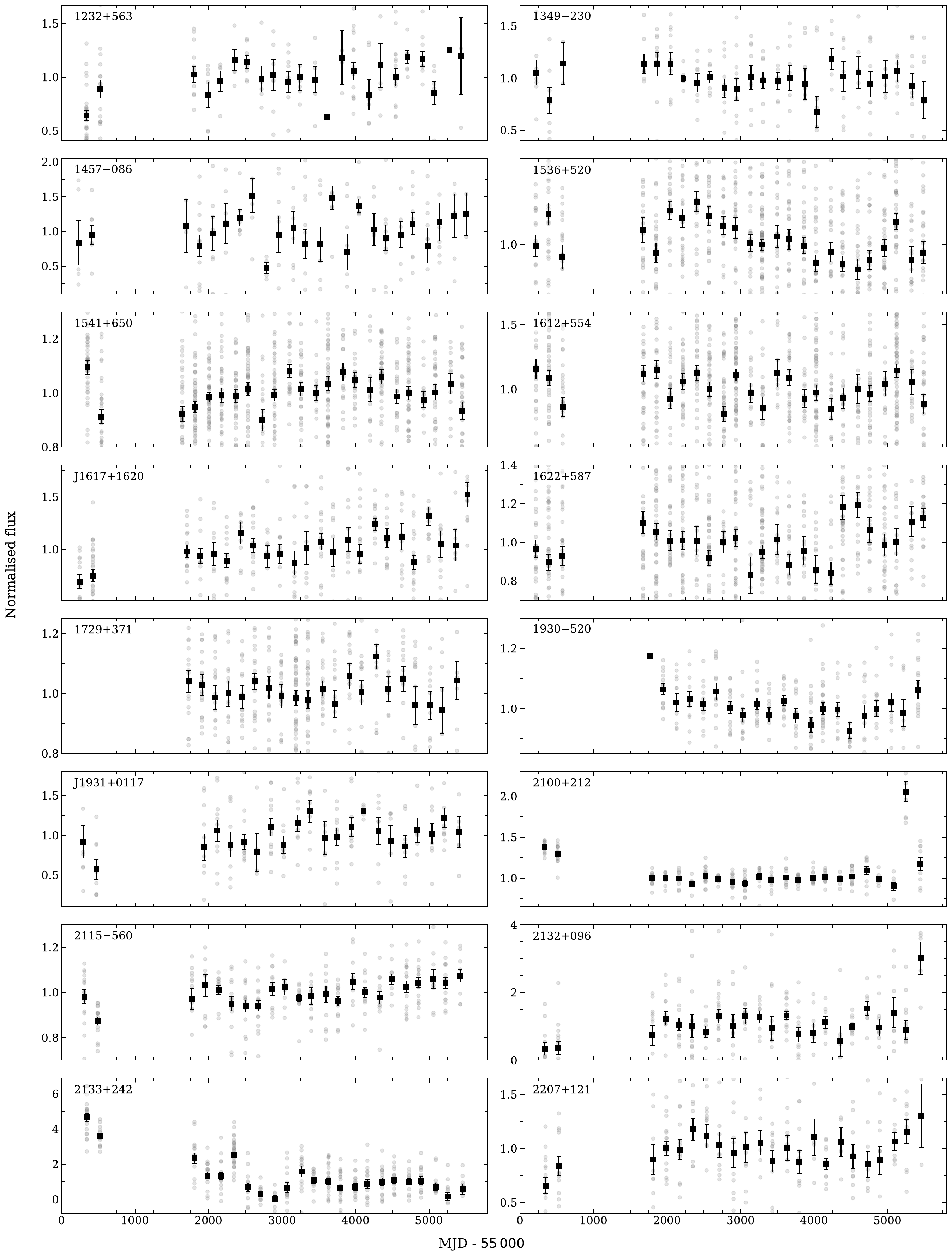}
\vskip -.2cm
\caption{ Cont.}
\end{figure*}

\addtocounter{figure}{-1}
\begin{figure*}
\vskip .1cm
\includegraphics[width=\textwidth]{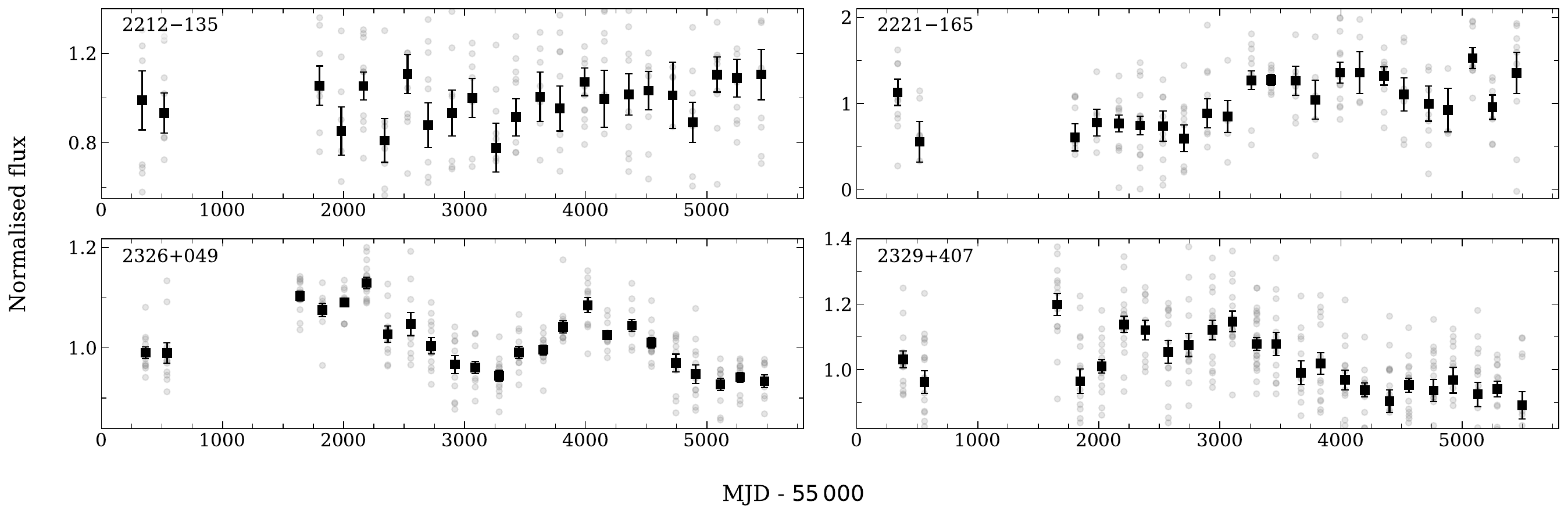}
\vskip -.2cm
\caption{ Cont.}
\end{figure*}

\addtocounter{figure}{-1}
\begin{figure*}
\vskip .1cm
\includegraphics[width=\textwidth]{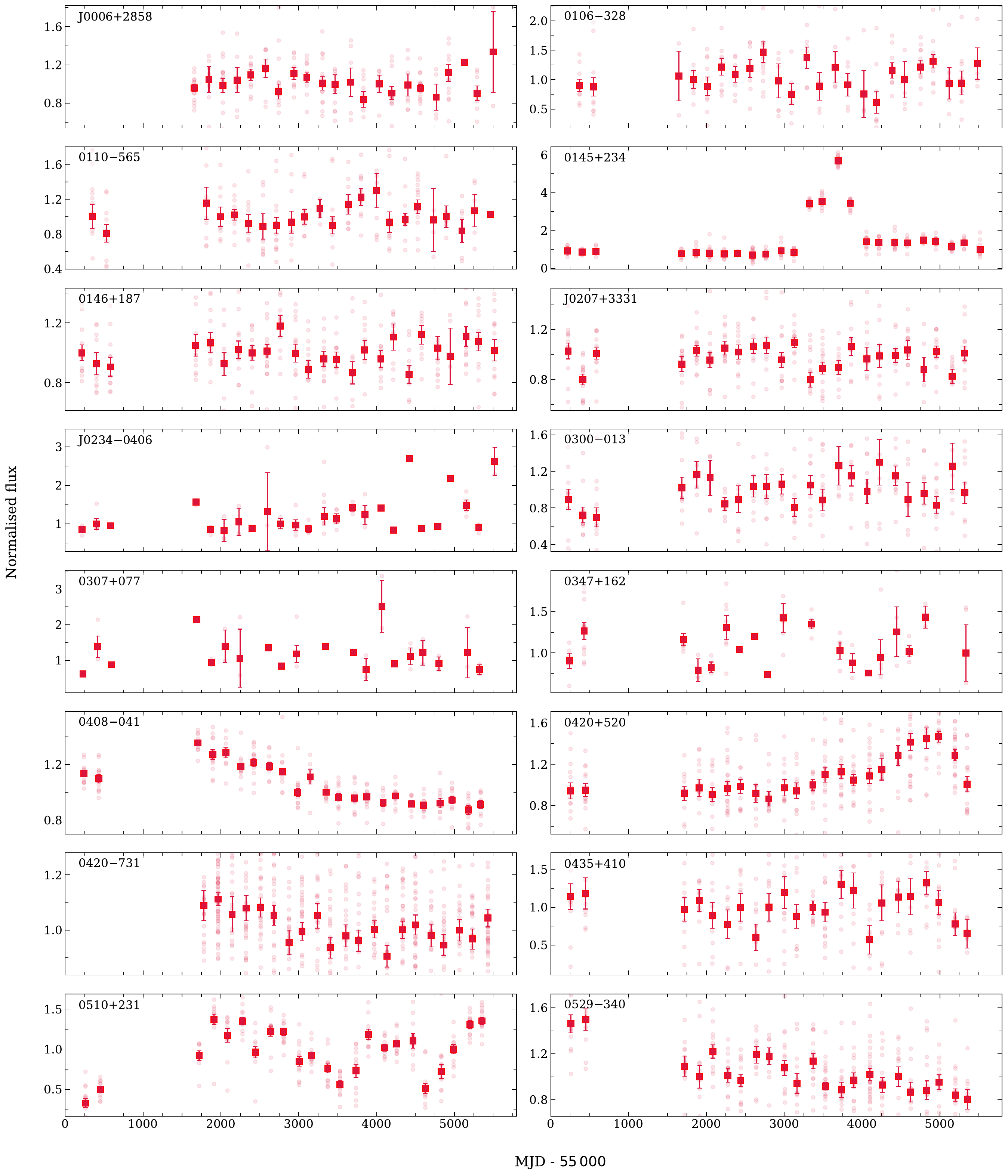}
\vskip -.2cm
\caption{ Cont.}
\end{figure*}

\addtocounter{figure}{-1}
\begin{figure*}
\vskip .1cm
\includegraphics[width=\textwidth]{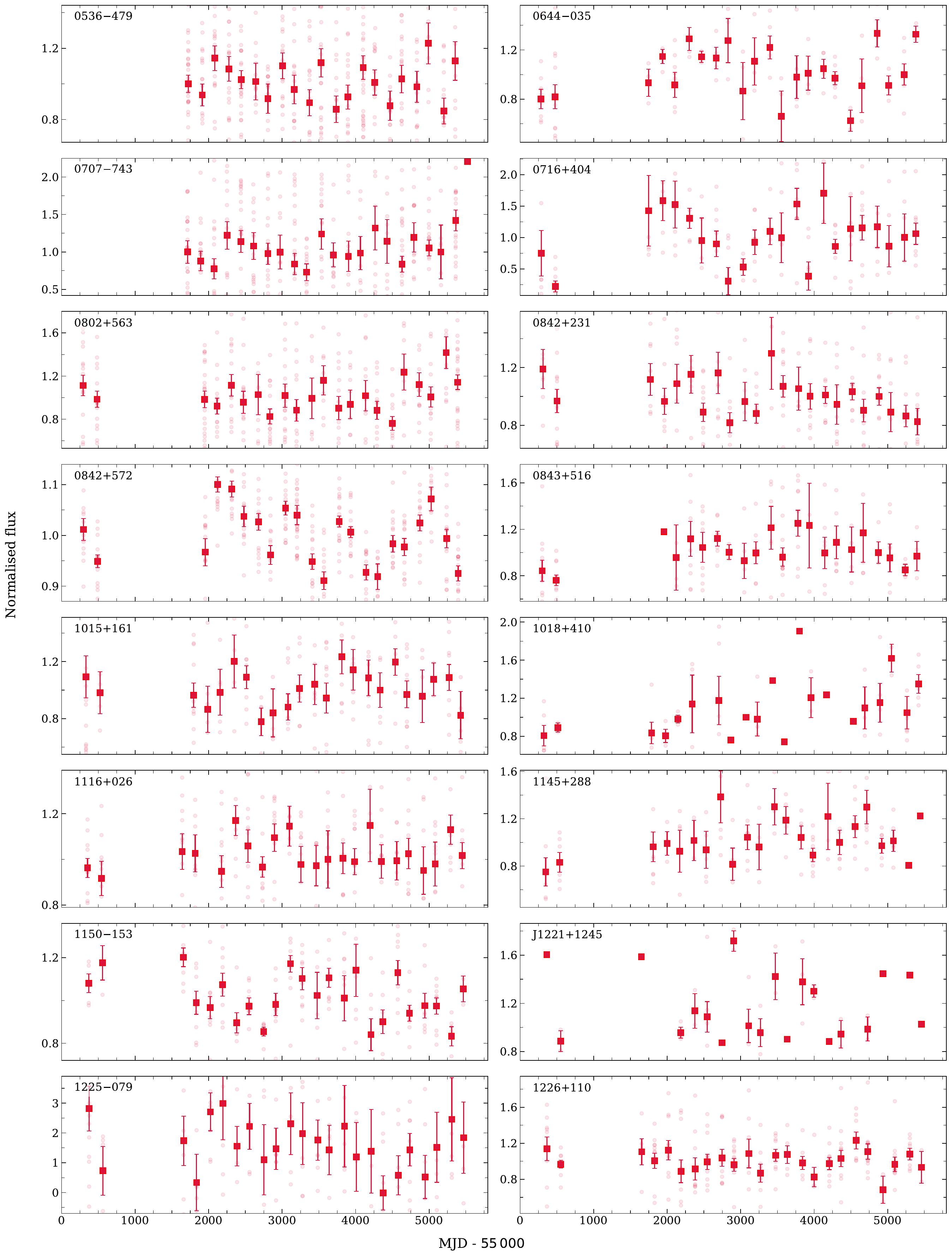}
\vskip -.2cm
\caption{ Cont.}
\end{figure*}

\addtocounter{figure}{-1}
\begin{figure*}
\vskip .1cm
\includegraphics[width=\textwidth]{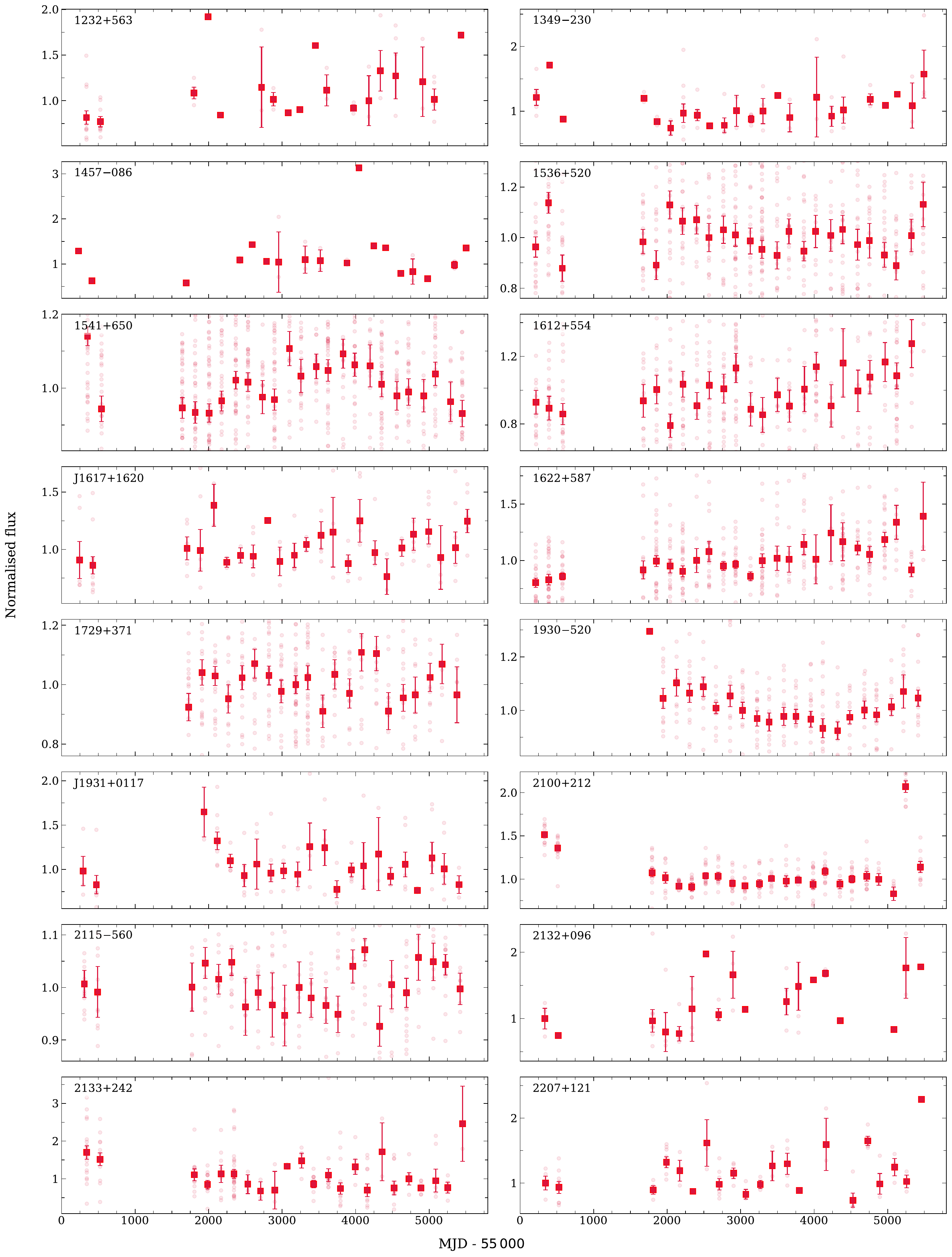}
\vskip -.2cm
\caption{ Cont.}
\end{figure*}

\addtocounter{figure}{-1}
\begin{figure*}
\vskip .1cm
\includegraphics[width=\textwidth]{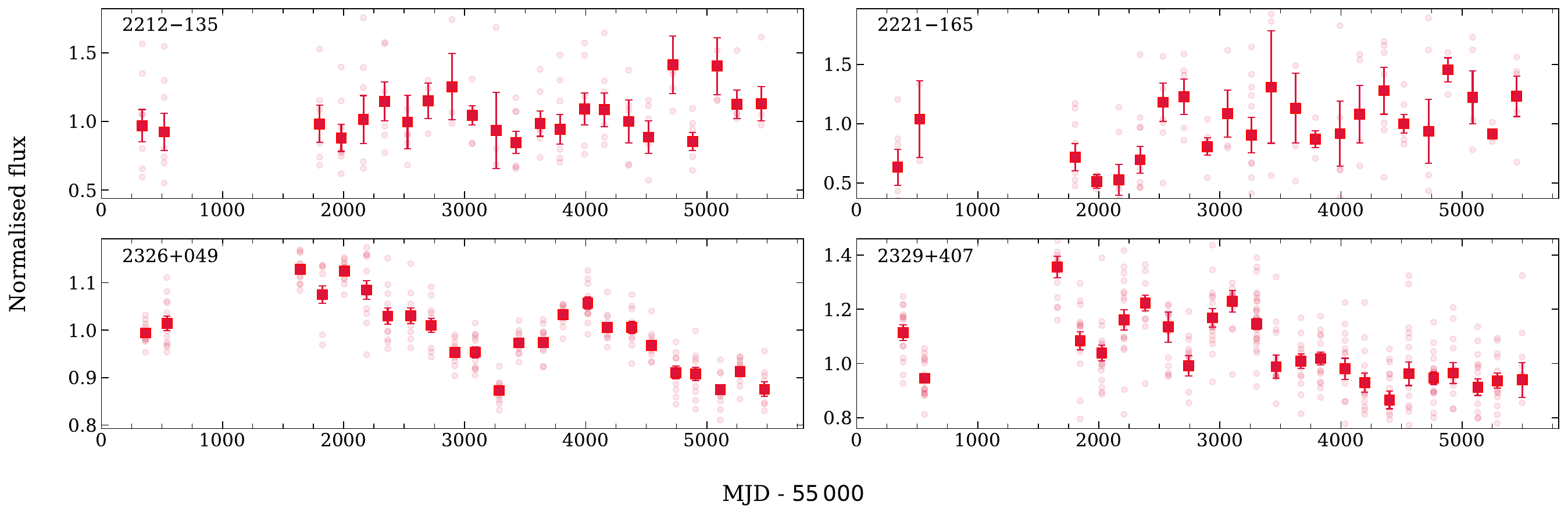}
\vskip -.2cm
\caption{ Cont.}
\end{figure*}

\bsp	
\label{lastpage}
\end{document}